%% file: main.tex
\documentclass[twocolumn,superscriptaddress,nofootinbib,floatfix,longbibliography]{revtex4-2}

\usepackage{amsmath} 					
\usepackage{amsfonts}
\usepackage{amssymb} 	
\usepackage{amsthm}
\usepackage{array}       				
\usepackage{graphicx}  					
\usepackage[usenames,dvipsnames]{xcolor} 
\usepackage{makeidx}   					
\usepackage[utf8]{inputenc} 			
\usepackage{url}     
\usepackage{times}

\usepackage[english]{babel}
	
\usepackage{siunitx}
\usepackage{booktabs}
\usepackage{csquotes}
\usepackage[shortlabels]{enumitem}
\usepackage{braket}
\usepackage{tcolorbox}
\usepackage{mathtools}
\usepackage{tikz}
\usetikzlibrary{quantikz}
\usetikzlibrary{calc}
\pgfdeclarelayer{bg}
\pgfsetlayers{bg,main}

\usepackage{hyperref}

\newtheorem{theorem}{Theorem}

\newcommand{\sgn}{\ensuremath{\operatorname{sgn}}}
\providecommand{\abs}[1]{\lvert #1 \rvert}
\newcommand{\Norm}[1]{\left\lVert #1 \right\rVert}
\newcommand{\dyad}[2]{|#1\rangle \! \langle #2|}

\newcommand{\mitigation}[1]{\mathsf{M\text{-}#1}}
\newcommand{\regularization}[1]{\mathsf{R\text{-}#1}}
\newcommand{\opId}{\operatorname{Id}}

\DeclareMathOperator{\A}{A} % Alignment
\DeclareMathOperator{\TA}{TA} % Target alignment
\newcommand{\transp}{\ensuremath{\scriptscriptstyle T}}
\newcommand*{\Tr}{\operatorname{Tr}}
\newcommand{\myvec}[1]{\ensuremath{\boldsymbol{#1}}}
\renewcommand{\aa}{\ensuremath{\myvec{a}}}
\newcommand{\bb}{\ensuremath{\myvec{b}}}
\providecommand{\ww}{\ensuremath{\myvec{w}}}
\providecommand{\xx}{\ensuremath{\myvec{x}}}
\providecommand{\yy}{\ensuremath{\myvec{y}}}
\providecommand{\ttheta}{\ensuremath{\myvec{\theta}}}
\providecommand{\calD}{\ensuremath{\mathcal{D}}}
\providecommand{\calN}{\ensuremath{\mathcal{N}}}
\providecommand{\calO}{\ensuremath{\mathcal{O}}}
\providecommand{\calS}{\ensuremath{\mathcal{S}}}
\providecommand{\calV}{\ensuremath{\mathcal{V}}}
\newcommand{\R}{\mathbb{R}}
\providecommand{\bbE}{\ensuremath{\mathbb{E}}}

\usepackage[version=2]{acro}
\DeclareAcronym{QEK}{short=QEK, long=quantum embedding kernel, first-long-format=\emph}
\DeclareAcronym{RBF}{short=RBF, long={radial basis function}, first-long-format=\emph}
\DeclareAcronym{SVM}{short=SVM, long=support vector machine, first-long-format=\emph}
\DeclareAcronym{QML}{short=QML, long=quantum machine learning, first-long-format=\emph}
\DeclareAcronym{SDP}{short=SDP, long=semi-definite program, first-long-format=\emph}
\DeclareAcronym{PQC}{short=PQC, long=parametrized quantum circuit, first-long-format=\emph}
\DeclareAcronym{QPU}{short=QPU, long=quantum processing unit, first-long-format=\emph}
\DeclareAcronym{iid}{short=i.i.d, long=independent and identically distributed, first-long-format=\emph}
\DeclareAcronym{NISQ}{short={NISQ}, long={noisy intermediate-scale quantum}, first-long-format=\emph} %
\DeclareAcronym{ML}{short=ML, long=machine learning, first-long-format=\text}

\begin{document}

\title{Training Quantum Embedding Kernels on Near-Term Quantum Computers}

\author{Thomas~Hubregtsen}
\email{thubregtsen@zedat.fu-berlin.de}
\affiliation{Dahlem Center for Complex Quantum Systems, Freie Universit\"{a}t Berlin, 14195 Berlin, Germany}

\author{David~Wierichs}
\affiliation{Institute for Theoretical Physics, University of Cologne, 50937 Cologne, Germany}

\author{Elies~Gil-Fuster}
\affiliation{Dahlem Center for Complex Quantum Systems, Freie Universit\"{a}t Berlin, 14195 Berlin, Germany}

\author{Peter-Jan~H.~S.~Derks}
\affiliation{Dahlem Center for Complex Quantum Systems, Freie Universit\"{a}t Berlin, 14195 Berlin, Germany}

\author{Paul~K.~Faehrmann}
\affiliation{Dahlem Center for Complex Quantum Systems, Freie Universit\"{a}t Berlin, 14195 Berlin, Germany}

\author{Johannes~Jakob~Meyer}
\affiliation{Dahlem Center for Complex Quantum Systems, Freie Universit\"{a}t Berlin, 14195 Berlin, Germany}
\affiliation{QMATH, Department of Mathematical Sciences, University of Copenhagen, 2100 Copenhagen, Denmark}

\date{\today}

\begin{abstract}
Kernel methods are a cornerstone of classical machine learning. The idea of using quantum computers to compute kernels has recently attracted attention.
\emph{Quantum embedding kernels (QEKs)} constructed by embedding data into the Hilbert space of a quantum computer are a particular quantum kernel technique that allows to gather insights into learning problems and that are particularly suitable for noisy intermediate-scale quantum devices.
In this work, we first provide an accessible introduction to quantum embedding kernels and then analyze the practical issues arising when realizing them on a noisy near-term quantum computer. 
We focus on quantum embedding kernels with variational parameters. These variational parameters are optimized for a given dataset by increasing the kernel-target alignment, a heuristic connected to the achievable classification accuracy.
We further show under which conditions noise from device imperfections influences the predicted kernel and provide a strategy to mitigate these detrimental effects which is tailored to quantum embedding kernels. We also address the influence of finite sampling and derive bounds that put guarantees on the quality of the kernel matrix.
We illustrate our findings by numerical experiments and tests on actual hardware.
\end{abstract}
\acresetall
\maketitle

\section{Introduction}
Quantum computing promises to solve problems that are currently intractable by exploiting the quantum nature of information. While long considered more a dream than a possibility, recent efforts have succeeded in constructing quantum devices able to perform computations intractable for classical computers~\cite{arute2019quantum}. This generation of quantum devices is referred to as \ac{NISQ} devices. Exploiting the non-classical capabilities of \ac{NISQ} devices to solve practically relevant problems is a rapidly growing field of research~\cite{cerezo2020variational,bharti2021noisy}.

\Ac{ML} on the other hand promises to leverage classical computers to solve ever more complicated problems. Especially the combination of deep neural networks and big data has led to impressive successes recently~\cite{vinyals2019grandmaster,brown2020language,senior2020improved}.

There exists, however, a variety of different approaches to construct learning models that are currently outshone by the more popular deep neural networks. Among those, \emph{kernel methods} are of particular interest as they provide a way to realize machine learning models that come with strong guarantees on their performance and offer a deep theoretical understanding that is often lacking when dealing with deep neural networks~\cite{scholkopf2018learning}.

It is no surprise that the idea of using near-term quantum computers for machine learning -- dubbed \ac{QML} -- has gained considerable traction lately~\cite{schuld2015introduction,biamonte2017quantum,mangini2021quantum}. The most prominent approach to construct learning models using \ac{NISQ} devices relies on the use of \acp{PQC}~\cite{schuldSupervisedLearningQuantum2018a,wittek2014quantum,lloydQuantumPrincipalComponent2014, schuldEffectDataEncoding2020}.
Kernel methods in particular have emerged as one particular candidate to realize \ac{QML} models~\cite{havlicekSupervisedLearningQuantum2019,schuld2019quantum_feature_hilbert_space,kusumoto2019experimental,bartkiewicz2020ExperimentalKernelFinite,blank2020quantum,huang2021power,peters2021machine}.
Furthermore, it was recently shown that other types of variational quantum learning models are fundamentally related to quantum kernel methods~\cite{schuldQuantumMachineLearning2021} and that quantum kernels enable the construction of learning problems that prove a separation between classical and quantum machine learning~\cite{liu2020rigorous,huang2021power}.
This work focuses specifically on \acp{QEK}, a subclass of quantum kernel methods where a \ac{PQC} is used to embed datapoints into the Hilbert space of quantum states of the underlying \ac{NISQ} device.

\begin{figure}
    \centering
    \includegraphics{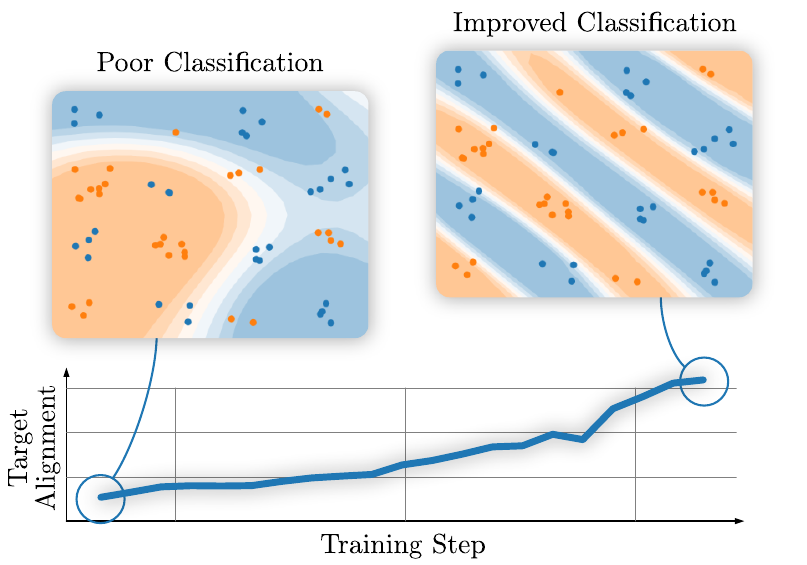}
    \caption{Quantum embedding kernels allow for classification of datasets through the use of a support vector machine. The quality of classification can be significantly improved by optimizing the parameters of the quantum embedding to increase the kernel-target alignment introduced in Sec.~\ref{sec:training_quantum_embedding_kernels}.}
    \label{fig:target_alignment}
\end{figure} 

\Acp{QEK} have certain appealing properties that make them attractive for use on \ac{NISQ} devices. As these devices have low coherence times, only quantum circuits of limited depth can be executed. As the computation of the kernel is only a small subroutine in a more general \ac{ML} technique, it does usually not require long coherence times. Furthermore, the coherence time of the \ac{NISQ} device can be exploited as much as possible by tailoring the \ac{PQC} realizing the \ac{QEK} to the underlying hardware. Another strong point is that noisy \acp{PQC} still lead to well-defined \acp{QEK} as will be explained in more detail later.
Kernel methods bring with them one main disadvantage: just constructing the kernel matrix already entails quadratic computational complexity in the number of training samples.
\Acp{QEK} are, ultimately, a particular case of kernels, and thus cannot but inherit this scaling, which hinders their application in contexts of large datasets. It is also often hard to choose the \emph{right} kernel for the problem at hand.

This work formalizes and extends our submission to the QHACK 2021 hackathon~\cite{hubregtsen2021code} hosted by Xanadu. We take an educational approach and analyze the whole pipeline necessary to make use of \acp{QEK}, including post-processing, error mitigation and optimization of variational parameters. We hope that this work can serve as accessible introduction for interested readers less familiar with \ac{QML} on near-term devices.  

Due to the educational nature of this work, it includes both pre-existing and novel contributions. We propose to use the \emph{kernel-target alignment} as a cost function to train the parameters of the \ac{QEK} to increase its performance on a particular dataset, as shown in Fig.~\ref{fig:target_alignment}. This technique is closely related to the \emph{metric quantum learning} approach proposed in Ref.~\cite{lloyd2020QuantumEmbeddingsML} for the purpose of optimizing \acp{PQC} for the encoding of a specific dataset.
Additionally, we discuss the inevitable noise from the underlying device, proposing a mitigation strategy tailored for \acp{QEK} that exploits the kernel's definition to infer the underlying noise levels.
We also analyze the influence of finite sampling on kernel quality -- an issue touched upon in Ref.~\cite{peters2021machine} -- showing that the number of samples typically required to achieve an approximation of fixed precision is of third order in the number of datapoints. We review pre-existing strategies that can be used to alleviate the influence of noise on the kernel matrix and propose a different one based on a semi-definite program. We close by providing numerical evidence that kernel training increases classification accuracy and that the proposed noise mitigation methods improve the quality of the obtained kernel matrix.

The rest of this work is organized as follows:
In Sec.~\ref{sec:kernels} and \ref{sec:quantum_embedding_kernels}, we give an intuitive introduction to the theory of kernels in general and quantum embedding kernels in particular.
Sec.~\ref{sec:training_quantum_embedding_kernels} introduces the kernel-target alignment as a measure of fit between dataset and kernel and motivates its use as a cost function for training quantum embedding kernels. In Sec.~\ref{sec:the_effects_of_noise}, we describe the influence of noise that effects the calculation of kernel matrices in a realistic setting and show how it can be mitigated using knowledge about the ideal kernel matrix. We additionally analyze the influence of finite sampling noise and show how regularization techniques can be used to still ensure that the kernel matrix stays positive semidefinite even. A pipeline for working with \acp{QEK} is suggested in Sec.~\ref{sec:qek_pipeline}. We present results of simulation obtained from both classical and quantum hardware that showcase the proposed approaches in Sec.~\ref{sec:numerical_experiments}.
We conclude by summarizing and discussing both our results and outstanding questions in Sec.~\ref{sec:summary_and_outlook}.

\section{Kernel Methods}\label{sec:kernels}
Kernel methods are among the cornerstones of \acl{ML}.
To understand what a kernel method does, let us first revisit one of the simplest methods to assign binary labels to datapoints: \emph{linear classification}.

\begin{figure}
    \centering
    \includegraphics{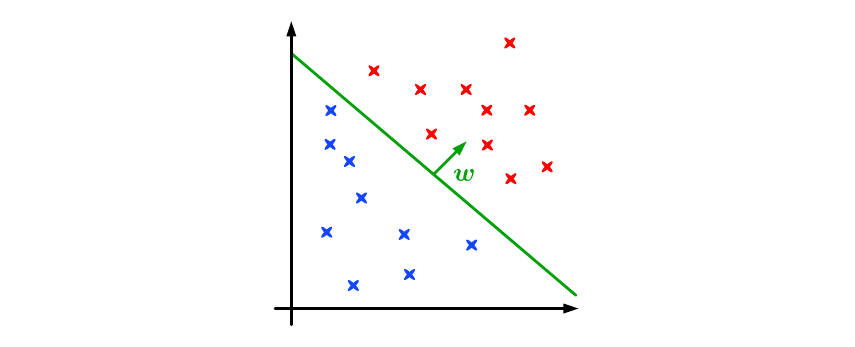}
    \caption{In linear classification, a line (or in higher dimensions a hyperplane) that separates the two classes is sought. We can define the tilt of the line via a vector $\ww$ orthogonal to it.}
    \label{fig:linear_classification}
\end{figure}

Imagine a set of points that lie in different parts of a plane. We want to construct a classifier that successfully predicts the classes of the datapoints. A linear classifier corresponds to drawing a line and assigning different labels $y = \pm 1$ representing the two classes to the opposing sides of the line, as depicted in Fig.~\ref{fig:linear_classification}.
Mathematically, this notion can be formalized by introducing a vector $\ww$ orthogonal to the line, thus determining its direction. We can then assign the class label $y$ to a datapoint $\xx$ via
\begin{align}\label{eq:lin_prediction}
   y(\xx) = \sgn(\langle \ww, \xx\rangle + b),
\end{align}
where the intercept term $b$ specifies the position of the line in the plane.
The same works for higher dimensional spaces too, where the vector $\ww$ does not just define a line, but a hyperplane.
It is immediately clear that this method is not very powerful, as datasets that are not separable by a hyperplane cannot be classified with high accuracy using this scheme. 

\begin{figure}
    \centering
    \includegraphics[width=0.48\textwidth]{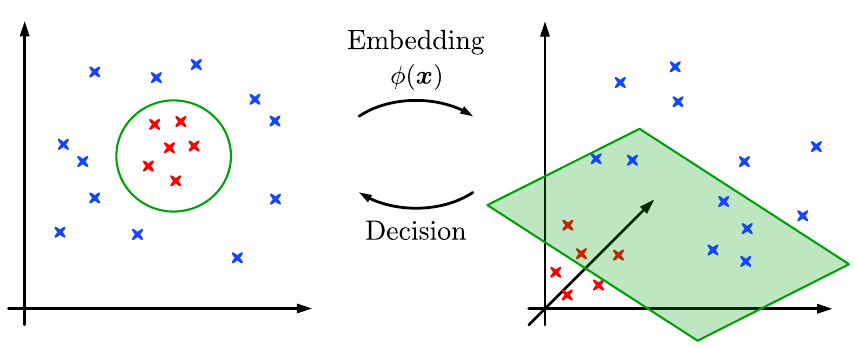}
    \caption{A non-linear embedding can be used to enhance the capabilities of a linear classifier. Linear classification in the embedding space can realize non-linear decision boundaries in the original space.}
    \label{fig:embedding_nonlinear_classification}
\end{figure}

There is, however, an ingenious way to enhance the capabilities of a linear classifier: One can specify a \emph{feature map} $\phi(\xx)$ that takes datapoints and embeds them into a larger \emph{feature space} and then perform linear classification in this feature space. In doing so, we can actually realize non-linear classification in the original space of our datapoints. This strategy is visualized in Fig.~\ref{fig:embedding_nonlinear_classification}.
We can modify the linear classifier of Eq.~\eqref{eq:lin_prediction} to include the feature map:
\begin{align}\label{eq:linear_classification_in_feature_space}
y(\xx) = \sgn(\langle \ww', \phi(\xx)\rangle + b),
\end{align}
where $\ww'$ lives in the feature space corresponding to the feature map $\phi$.

A major result in kernel theory is the \emph{representer theorem}~\cite{scholkopf2002learning}. It states that one can write the vector $\ww'$ that defines an optimal decision boundary as a sum of the embedded datapoints with real coefficients: $\ww' = \sum_i \alpha_i \phi(\xx_i)$\footnote{The representer theorem makes mild assumptions about the way we measure \enquote{optimal}, but for our applications these are always fulfilled.}. Inserting this into Eq.~\eqref{eq:linear_classification_in_feature_space} yields
\begin{align}\label{eq:linear_kernel_prediction}
    y(\xx) = \operatorname{sgn}\left(\sum_i \alpha_i \langle \phi(\xx_i), \phi(\xx)\rangle + b\right).
\end{align}
While this might not seem useful at first, notice that the above formula only contains inner products between vectors in the embedding space,
\begin{align}
    k(\xx, \xx') = \langle \phi(\xx), \phi(\xx')\rangle.
\end{align}
We call the function $k$ the \emph{kernel} associated to the feature map $\phi$. But why do kernels deserve all the attention they get?

The relevant insight is that we can often find an explicit formula for the kernel $k$ without explicitly performing the feature map $\phi$.
Consider for example the embedding
\begin{align}
    \phi\colon\begin{pmatrix}x_1 \\ x_2 \end{pmatrix} &\mapsto \begin{pmatrix} x_1^2 \\ \sqrt{2} x_1 x_2 \\ x_2^2 \end{pmatrix},
\end{align}
whose associated kernel can be explicitly calculated:
\begin{align}
k(\xx, \xx')
&= \langle \phi\begin{pmatrix}x_1 \\ x_2 \end{pmatrix}, \phi\begin{pmatrix}x'_1 \\ x'_2 \end{pmatrix}\rangle \\
&= x_1^2 {x'}_1^2 + 2 x_1 x_2 x'_1 x'_2 + x_2^2 {x'}_2^2 \\
&= (x_1 x'_1 + x_2 x'_2)^2 \\
&= \langle \xx, \xx' \rangle^2.
\end{align}
We find that it can be obtained by simply computing the inner product of two vectors \emph{in the initial space} and squaring it. Implicitly, we are however computing the inner product relative to the embedding $\phi$, i.e., in feature space! 
This is the central property of kernel-based methods. In many relevant cases the embedding will require a much higher cost to compute than the kernel, while one still gains access to the larger feature space through the kernel. This implicit use of the embedding through its associated kernel is known as the \emph{kernel trick}.

If we do not need the embedding at all, how can we then determine if a given function $k$ is actually a kernel for some feature map? This question is answered by checking the \emph{Mercer condition}, which states that any function fulfilling
\begin{align}
    \sum_{i,j} c_i c_j k(\xx_i, \xx_j) \geq 0
\end{align}
for \emph{all} possible sets of real coefficients $\{c_i \}$ and sets of datapoints $\{ \xx_i \}$ is a kernel. Alternatively, we can check whether the kernel matrix $K$ with entries
\begin{align}\label{eq:kernel_matrix}
    K_{ij} = k(\xx_i, \xx_j)
\end{align}
associated with \emph{any} dataset $\{ \xx_i \}$ is always \emph{positive semi-definite}.

If we now come back to the example of linear classification in a feature space that motivated our introduction to kernel methods, it is natural to ask how we can best choose the separating hyperplane.
The most common strategy and application for kernel methods is the \ac{SVM}.
The idea behind \acp{SVM} is to find the hyperplane with the maximal \emph{margin}.
The margin describes the distance of the dataset on either side of the hyperplane.
Intuitively, a larger margin is better, since the result would be that outliers of the dataset have a smaller chance of being wrongly classified.
This way, \ac{SVM} is an algorithm that takes as input the kernel matrix from Eq.~\eqref{eq:kernel_matrix} and delivers the values $\alpha_i$ and $b$ for Eq.~\eqref{eq:linear_kernel_prediction} that correspond to the decision boundary with the maximal margin.

To predict a class label for a new datapoint, we need to calculate the kernel with respect to the training datapoints and decide on a class label, as shown in Eq.~\eqref{eq:linear_kernel_prediction}. A strong advantage of \acp{SVM} is that usually only few weights $\{\alpha_i \}$ contribute significantly to the sum in Eq.~\eqref{eq:linear_kernel_prediction}. We can thus make a prediction by calculating the kernel with respect to these datapoints from the training dataset. The corresponding datapoints are the eponymous \emph{support vectors} -- as they support the decision boundary. Intuitively, we can imagine that comparing a new datapoint only to points close to the decision boundary will give important information about the class label.

Kernel methods are not limited to classification. In fact, one can take any \ac{ML} technique that can be reformulated in terms of inner products and replace the inner products by kernel functions to get a \enquote{kernelized} variant. This leads to interesting applications such as kernel principal component analysis~\cite{goos1997kernel} or kernel ridge regression~\cite{saunders1998ridge}.

While we have now seen that kernel methods can enhance the power of many \ac{ML} techniques, the current progress of learning models is driven by deep neural networks, not kernel methods. This is due to their following downsides: For the application of kernel methods, the kernel matrix with respect to the input data needs to be constructed -- which has quadratic complexity in the number of datapoints. This can already constitute a substantial impediment in the world of big data where the number of datapoints can be in the millions.
Another downside is that the selection of a suitable kernel for a given problem is a non-trivial task. The \emph{\ac{RBF} kernel} also known as \emph{Gaussian kernel} given by
\begin{align}\label{eq:def_rbf_kernel}
    k_{\text{RBF}}(\xx, \xx') = \exp\left(-\frac{\lVert \xx - \xx' \rVert^2}{2 \sigma^2} \right)
\end{align}
is often a decent starting point, but even there the parameter $\sigma$ that quantifies how close datapoints need to be to be considered similar needs to be fine-tuned.

Despite the fact that kernel methods do not dominate contemporary \ac{ML} applications, they are extremely useful to understand learning models in general. This is because many learning methods can be mapped back to kernel methods for which in turn there exist theoretical guarantees, for example on their capability to generalize on unseen data~\cite{scholkopf2002learning}.
    
\section{Quantum Embedding Kernels}\label{sec:quantum_embedding_kernels}

On \ac{NISQ} hardware, we make use of quantum gates, like Pauli rotations, to load data onto the quantum computer. This constitutes a quantum circuit that is represented by a unitary operation dependent on the specific datapoint, $U(\xx)$. As soon as the data is loaded, the quantum system is in a state that depends on the datapoint in question, 
\begin{align}
    \ket{\phi(\xx)} = U(\xx) \ket{0}.
\end{align}
This approach is also known as a \emph{quantum feature map}~\cite{schuld2019quantum_feature_hilbert_space} because we are effectively embedding the datapoint in the Hilbert space of quantum states.
As we learned in Sec.~\ref{sec:kernels}, it is no large step from a feature map to a kernel function. We only need to take the inner product between quantum states, which is given by the \emph{overlap}
\begin{align}\label{eq:qek_pure_state}
    k(\xx,\xx') &= \abs{\braket{\phi(\xx')|\phi(\xx)}}^2.
\end{align}
This is the \emph{\acf{QEK}} associated to the quantum feature map $\ket{\phi(\xx)}$.

In general, we are not able to avoid noise, which means that we cannot assume that the embedded quantum state is pure. In this case, the quantum embedding is realized by a data-dependent density matrix $\rho(\xx)$, which for a pure state reduces to $\rho(\xx) = \dyad{\phi(\xx)}{\phi(\xx)}$. The inner product is given by
\begin{align}\label{eq:qek_mixed_state}
    k(\xx,\xx') = \Tr\{\rho(\xx)\rho(\xx')\}.
\end{align}
This inner product is also known as the Hilbert-Schmidt inner product for matrices. For pure quantum states, this reduces to Eq.~\eqref{eq:qek_pure_state}.

In summary, any quantum feature map induces a \ac{QEK}. We can use this kernel as a subroutine in a classical kernel method, for example the \ac{SVM}, which yields a hybrid quantum-classical approach. In this case, the separating hyperplane is constructed in a purely classical manner, only the kernel function between the training datapoints is evaluated on the quantum computer. 

To be able to use \acp{QEK} in this way, we need to evaluate the overlap of two quantum states on near-term hardware. There are a number of advanced algorithms to estimate the overlap of two quantum states ~\cite{fanizzaSwapTestOptimal2020,buhrman2001QuantumFingerprinting,Cincio2018LearningAlgorithmStateOverlap,huang2020predicting,flammiaDirectFidelityEstimation2011}.
All these algorithms work for arbitrary states, and so they are agnostic to how the states were obtained by necessity.
By exploiting the structure and specifics of \acp{QEK}, though, we can do better.

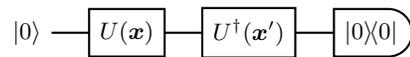
\begin{figure}
    \centering
    \begin{quantikz}
    \lstick{$\ket{0}$}&\gate{\vphantom{\int}U(\xx)}&\gate{\vphantom{\int}U^{\dagger}(\xx')}& \meterD{\vphantom{\int}\lvert 0\rangle\!\langle 0\rvert}
    \end{quantikz}  
    \caption{The overlap between the embedded states can be computed by applying the unitary $U(\xx)$ embedding the first datapoint and the adjoint of the unitary embedding the second datapoint $U^{\dagger}(\xx')$. This approach results in a doubled circuit depth but does not need auxiliary qubits. It works only for pure states.}
    \label{fig:adjoint}
\end{figure}

For unitary quantum embeddings, i.e.\ embeddings resulting in a pure quantum state, this is straightforward if we can construct the adjoint of the data-encoding circuit, $U^\dagger(\xx)$. In this case, we can rewrite the overlap as
\begin{align}\label{eq:adjoint}
    \abs{\braket{\phi(\xx')|\phi(\xx)}}^2 = \abs{\braket{0|U^\dagger(\xx')U(\xx)|0}}^2.
\end{align}
This is nothing but the probability of observing the $\ket{0}$ state when measuring the state $U^\dagger(\xx')U(\xx)|0\rangle$ in the computational basis.
In order to obtain an estimate, we can therefore initialize the quantum system in the $|0\rangle$ state, apply the unitary operation $U(\xx)$ followed by $U^\dagger(\xx')$, and finally measure in the computational basis.
From there, we only need to record the frequency with which the prepared state is found in the $\ket{0}$ state to obtain our estimate.
The circuit diagram for this \emph{adjoint} approach can be found in Fig.~\ref{fig:adjoint}. 
This scheme does not need auxiliary qubits, yet it applies the data-encoding circuit twice (or the adjoint thereof).
This way, while the width of the circuit for the adjoint approach does not increase from the data-encoding one, the depth is doubled.

\begin{figure}
    \centering
    \begin{quantikz}
    \lstick{$\ket{0}$}&\gate{H}&\ctrl{2}&\gate{H}&\meterD{Z}  \\
    \lstick{$\ket{0}$}&\gate{U(\xx)}&\swap{1}&\qw&\qw  \\
    \lstick{$\ket{0}$}&\gate{U(\xx')}&\targX{} &\qw&\qw
    \end{quantikz}  
    \caption{The overlap between the embedded states can be computed by embedding both datapoints in parallel. An auxiliary qubit is then used together with controlled SWAP operations to extract the overlap, which is obtained from the expectation value of the Pauli-$Z$ observable on the auxiliary qubit. This approach results in a doubled circuit width and requires one additional qubit and controlled SWAP operations. It works for pure and mixed states.}
    \label{fig:swap_test}
\end{figure}
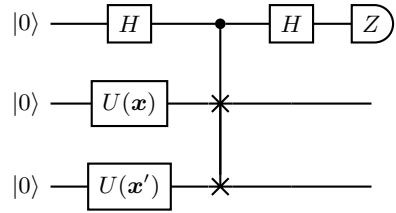

Another alternative approach to estimate the overlap between two quantum states is the \emph{SWAP test}. The SWAP test is based on the \emph{SWAP trick}, a mathematical gimmick that allows us to transform the product of the density matrices in Eq.~\eqref{eq:qek_mixed_state} into a tensor product:
\begin{align}
    k(\xx, \xx') &= \Tr\{\rho(\xx)\rho(\xx')\} \\
    &= \Tr \{ (\rho(\xx) \otimes \rho(\xx')) S \},
\end{align}
where $S$ denotes the SWAP operation between the two quantum systems in the states $\rho(\xx)$ and $\rho(\xx')$. To extract this quantity, the SWAP test makes use of an auxiliary qubit that controls the SWAP operation. The exact circuit is depicted in Fig.~\ref{fig:swap_test}. 
While this approach needs auxiliary qubits and a quantum computer roughly twice as wide, its depth increases only ever so slightly, and it also works for mixed quantum states.
If the deeper requirements of the adjoint approach were too limiting, the SWAP test would be a natural alternative.

As quantum feature maps occur in many applications of \ac{NISQ} computing, it is no surprise that \acp{QEK} are instrumental beyond their use as subroutines for classical machine learning methods. In Ref.~\cite{schuldQuantumMachineLearning2021} it was shown that all variational quantum learning methods boil down to kernel methods, providing an opening for kernel theory to explore the properties of these learning models.

\section{Training Quantum Embedding Kernels}
\label{sec:training_quantum_embedding_kernels}
Quantum feature maps can have variational parameters. In this section, we discuss how to adjust these parameters to improve the classification capabilities of quantum embedding kernels.

Before we can dive into the details we take a step back to clear up the terminology we use in this section. Adjusting the variational parameters of a quantum feature map -- and therefore of its associated \ac{QEK} -- can be seen as a particular instance of \emph{model selection}. For our purposes, we again split this into two steps, first \emph{kernel selection} and second \emph{kernel optimization}. 

Kernel selection is choosing a particular quantum circuit \emph{layout}, or \emph{ansatz}, with certain variational parameters. 
Here, we refer to a \emph{variational family} of kernels, much like the family of \ac{RBF} kernels defined in Eq.~\eqref{eq:def_rbf_kernel} where we vary the standard deviation $\sigma$. 
Kernel optimization corresponds to the process of fixing the variational parameters to ensure a good classification performance on our target data.
Both steps are important to assure a good performance of kernel-based methods. In the following, we will shed light on the possible ways to perform kernel optimization.

A straightforward way to perform kernel optimization is \emph{exhaustive parameter search}.
For a single parameter $\theta\in\R$ this method consists of fixing a range of parameter values $[\theta_{\min},\theta_{\max}]$ and an $\varepsilon$-cover of it:
\begin{align}
    \{\theta_{\min}, \theta_{\min}+\varepsilon, \ldots, \theta_{\max}-\varepsilon, \theta_{\max}\}.
\end{align}
Then, a \ac{SVM} is fitted to the training data for every value in the $\varepsilon$-net and we keep the one which attains the highest accuracy score.
This way, we need to fit roughly $p:=(\theta_{\max}-\theta_{\min})/\varepsilon$ many \acp{SVM} in order to pin down the optimal choice of $\theta$.
If, on the contrary, our kernel comprised a vector of parameters $\ttheta\in\R^r$, we would still proceed in an analogous way:
In a simplified case, we would still fix a range of parameter values $\ttheta\in[\theta_{\min},\theta_{\max}]^r$ and an $\varepsilon$-net.
Notice how, now, the $\varepsilon$-net would contain approximately $p^r$ sites, since each parameter component $\theta_i$ would need to take all possible values between $\theta_{\min}$ and $\theta_{\max}$ for every possible combination of the remaining parameter components!
It is thus clear to see how the computational complexity of exhaustive parameter search grows exponentially $\calO(\exp(r))$ with respect to the parameter vector dimension $r$.
One particular example of this type is the \enquote{leave-one-out} error bound~\cite{scholkopf2002learning}, where after fitting the \ac{SVM} to all points but one, the predicted label for the left out point describes the quality of the kernel.
Because of the unfavorable scaling, exhaustive parameter search procedures are thus only suitable for optimizing few parameters.

It is therefore sensible to resort to a proxy quantity that is easier to compute but still acts as a predictor for classification accuracy.
Ref.~\cite{wangOverviewKernelAlignment2015} provides an overview of such measures as provided in Refs.~\cite{cortes2012CenteredAlignment, baramLearningPolarization, wang2009LocalKernelPolarization, Wang2008KernelClassSeparability, xiong2005EmpiricalFeatureSpace, nguyen2008kernelmatrixevaluationmeasure}, using the \emph{kernel-target alignment} from Ref.~\cite{cristianiniKernelTargetAlignment2006} as central building block.

In the following, we assume initially that the training dataset is \emph{balanced}, meaning that it contains equally many datapoints per class. The generalization to unbalanced datasets is straightforward and left to the end of this section.

The central idea behind the kernel-target alignment is that the \emph{labels} for the training set can be seen as an instance of a very particular kernel, which acts as an oracle that always outputs the correct similarity for two datapoints:
\begin{align}
    k^{*}(\xx, \xx') = \begin{cases}
    1 &\text{if $\xx$ and $\xx'$ in same class}\\
    -1 &\text{if $\xx$ and $\xx'$ in different classes}
    \end{cases}
\end{align}
Of course, in general we do not have access to this \emph{ideal} kernel, but on the training data it is given by the training labels and the kernel matrix predicted by this ideal kernel has entries
\begin{align}
    K^{*}_{ij} = y_i y_j.
\end{align}
This means that, if we place the labels into a vector $\yy$, we can express the ideal kernel matrix as the outer product of that vector with itself:
\begin{align}
    K^{*} = \yy \yy^{\transp}.
\end{align}

To get to a measure of how well a kernel captures the nature of the training dataset we need a way to compare the kernel matrix with the ideal one. To obtain the kernel-target alignment we will make use of geometric reasoning. Remember that we can measure the \emph{alignment} of two vectors $\aa$ and $\bb$ by evaluating and normalizing the inner product:
\begin{align}
    \A(\aa, \bb) = \frac{\langle \aa, \bb \rangle}{\sqrt{\langle \aa, \aa \rangle \langle \bb, \bb \rangle}}.
\end{align}
The alignment is related to the angle $\sphericalangle(\aa, \bb)$ between the vectors as $\cos \sphericalangle(\aa, \bb) = \A(\aa, \bb)$, which means that the alignment is a quantity that ranges from -1 for vectors pointing in exactly opposite directions to +1 for vectors pointing in exactly the same direction.

We can apply the same reasoning to kernel matrices. To do so, we need to define an inner product between two matrices. For the definition of the kernel-target alignment we will use the \emph{Frobenius inner product}. For that, we simply treat the matrices as if they were column vectors, with every entry of the matrix being a separate entry of the vector. This means that the inner product is just
\begin{align}
    \langle A, B \rangle_F = \sum_{ij} A_{ij} B_{ij} = \Tr \{ A^{\transp} B \},
\end{align}
from where one defines the alignment of two matrices $B$ and $B'$ as
\begin{align}\label{eq:matrix_alignment}
    \A(B, B') = \frac{\langle B, B' \rangle_F}{\sqrt{\langle B, B \rangle_F \langle B', B' \rangle}_F}.
\end{align}

Now we have all the ingredients to define the kernel-target alignment:
\begin{align}\label{eq:kernel_target_alignment_kernel_matrix}
    \TA(K) &= \A(K, K^{*})= \frac{\langle K, K^{*} \rangle_F}{\sqrt{\langle K, K \rangle_F \langle K^{*}, K^{*} \rangle_F}}.
\end{align}
We can equivalently express this in terms of the kernel function and the training dataset: 
\begin{align}\label{eq:kernel_target_alignment_kernel_function}
    \TA(k) &= \frac{\sum_{ij} y_i y_j k(\xx_i, \xx_j)}{\sqrt{\left(\sum_{ij} k(\xx_i, \xx_j)^2\right)\left(\sum_{ij} y_i^2 y_j^2\right)}} \\
    &= \frac{\sum_{ij} y_i y_j k(\xx_i, \xx_j)}{n\sqrt{\sum_{ij} k(\xx_i, \xx_j)^2}},\label{eq:kernel_target_alignment_kernel_function_balanced}
\end{align}
where we used the fact that for all labels $y_i^2 = 1$ and $n$ denotes the number of points in the training set.

At the beginning of this section we assumed that the training set was balanced, i.e., that it contains the same number of datapoints for each class. If this were not the case, the approach we just outlined would run into problems,
because the contributions from one class would dominate the kernel-target alignment. We could however mitigate this by simply rescaling the labels, dividing them by the number of samples available in their class. In this case, we cannot use Eq.~\eqref{eq:kernel_target_alignment_kernel_function_balanced} but have to stay with Eq.~\eqref{eq:kernel_target_alignment_kernel_function}.

We can gather further intuition why the kernel-target alignment is a meaningful measure by looking at the numerator of Eq.~\eqref{eq:kernel_target_alignment_kernel_function}, $\sum_{ij} y_iy_j \, k(\xx_i,\xx_j)$. This quantity is also known as the \emph{kernel polarity}. Each term $y_iy_j \, k(\xx_i,\xx_j)$ in the sum is a product of the kernel function of two points and their labels.
If both points belong to the same class, $y_i y_j = +1$, the kernel value increases the kernel-target alignment, whereas if the labels are different $y_i y_j = -1$, the term decreases it.
Increasing the kernel-target alignment therefore means both increasing the kernel values for datapoints from the same class and decreasing them for datapoints in different classes.
\begin{figure}
    \centering
    \includegraphics{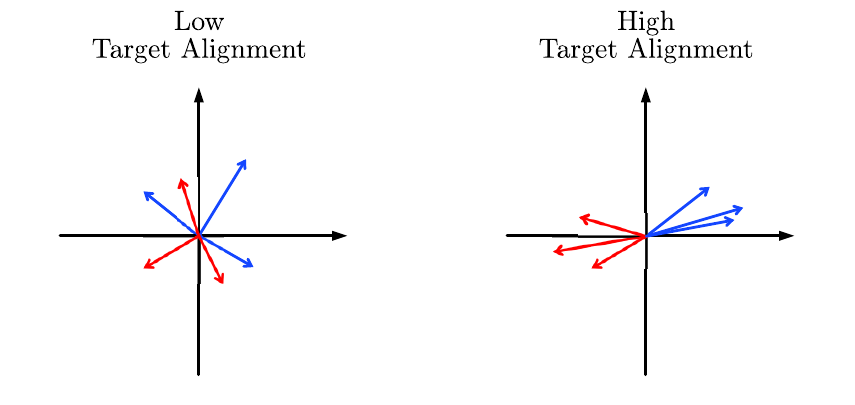}
    \caption{The kernel-target alignment is high if the feature vectors corresponding to datapoints in the same class cluster together and the feature vectors of points in the opposite class lie exactly opposite to them. This illustrates that a high kernel-target alignment allows for easier linear separability.}
    \label{fig:ta_illustration}
\end{figure}
Fig.~\ref{fig:ta_illustration} visually illustrates how a large kernel-target alignment allows for easier linear classification of the training data. Beyond this intuition, the kernel-target alignment profits from theoretical guarantees regarding both its high concentration about the expected value and its good generalization behavior in labeling previously unseen data~\cite{cristianiniKernelTargetAlignment2006, kandola2002optimizingalignmentcombinations,nguyen2008kernelmatrixevaluationmeasure}. 

With the kernel-target alignment we now have a measure that we can use as a cost function to maximize through a hybrid quantum-classical optimization loop~\cite{mcclean2016theory}. At every iteration, the \ac{QPU} is used to evaluate the kernel matrix $K$, recall constructing $K$ has quadratic complexity in the number of datapoints.

It turns out that optimizing the kernel-target alignment is closely related to the \enquote{quantum metric learning} approach put forward in Ref.~\cite{lloyd2020QuantumEmbeddingsML} that analyzed different strategies to optimize quantum feature maps. Indeed, the Hilbert-Schmidt distance-based method is the same as optimizing the unnormalized kernel-target alignment, the polarity. We detail the connection in App.~\ref{sec:app_feature_map_opt}.

\section{The Effects of Noise}
\label{sec:the_effects_of_noise}
Noise is one of the namesakes of \ac{NISQ} devices and considering the effects of noise is therefore of utmost importance. In this section, we discuss how both noise arising from imperfect quantum operations -- device noise -- and noise arising from finite sampling of expectation values affects \acp{QEK} and how it can be mitigated.

\subsection{Device Noise}\label{sec:device_noise}
\Ac{NISQ} devices suffer from unavoidable noise caused by unintentional interactions with the environment or imperfect control. 
It is thus not possible to prepare pure quantum states with an embedding circuit. 
This fact has multiple implications for \acp{QEK} and their implementation.

Noise can be modeled by \emph{quantum channels}. Formally, any map that takes valid quantum states to valid quantum states can be seen as a quantum channel. An example is \emph{depolarizing noise}, which corresponds to a complete loss of information about the underlying quantum state with a certain probability $1-\lambda$. We formally realize it by replacing the system's quantum state with the maximally mixed state with probability $1-\lambda$:
\begin{align}\label{eq:depolarizing_noise}
    \calD_{\lambda}[\rho] = \lambda \rho + (1-\lambda)\frac{1}{2^N}.
\end{align}
Depolarizing noise is a popular model for noise in quantum systems, as it is simple and subsumes other, more nuanced noise models.

As we have seen in Sec.~\ref{sec:quantum_embedding_kernels}, the \acp{QEK} can also be defined for mixed states for which it corresponds to the state overlap. The SWAP test can be directly employed to compute the overlap of mixed states (see Fig.~\ref{fig:swap_test}), but often we would like to use the adjoint method due to its lower qubit requirements (see Fig.~\ref{fig:adjoint}). The adjoint method, however, needs more consideration, because the implementation of the adjoint noisy embedding circuit itself is not straightforward. But if we fail to implement the correct adjoint operation, we are no longer computing the overlap of the quantum embedding states and therefore also do not compute a valid kernel!

In the noiseless embedding circuit, all operations are unitary and typically the adjoint of every elementary operation is available. This becomes apparent when we consider that any quantum circuit can be constructed from controlled NOT gates, which are self-adjoint, and single-qubit Pauli rotations, whose adjoint is obtained by performing the same rotation with negated angle.

The device noise, however, can in principle prevent us from implementing the adjoint embedding. As an example, consider a quantum channel that represents a noisy Pauli rotation gate, $\calV(\theta)$. We model it by the original rotation gate $R(\theta)$ that is followed by a noise channel $\calN$. The channel $\calN$ could model imprecision in the control of the rotation angle, unwanted interactions with the environment or other noise processes, but we will leave it arbitrary for the sake of the example. The noisy gate is then given by
\begin{align}
    \calV(\theta)[\rho] = \calN[R(\theta) \rho R^{\dagger}(\theta)].
\end{align}
The adjoint of the noisy Pauli rotation gate is then given by
\begin{align}
    \calV(\theta)^{\dagger}[\rho] = R(-\theta) \calN^\dagger[\rho ] R^{\dagger}(-\theta).
\end{align}
But how would we implement $\calV(\theta)^{\dagger}$? The very nature of a noise channel implies that we cannot control it or choose at which time the noise occurs. Instead, we have to work with the noisy quantum gates at our disposal, which means that we can only approximate $\calV(\theta)^{\dagger}$ by $\calV(-\theta)$:
\begin{align}
    \calV(-\theta)[\rho] = \calN[R(-\theta) \rho R^{\dagger}(-\theta)] \neq \calV(\theta)^{\dagger}.
\end{align}
In general, this approximation is not equal to the adjoint of the noisy unitary. This only happens if the noise channel $\calN$ is self-adjoint and commutes with the unitary operation $R(\theta)$, as is true for the depolarizing noise introduced in Eq.~\eqref{eq:depolarizing_noise}. 

Let us now take a step back and look at actual \ac{NISQ} devices. They are usually programmed at the gate-level, assuming perfect unitaries. The adjoint of a perfect unitary circuit is readily available, but only if the behavior of the available \ac{NISQ} device is well-modeled by depolarizing noise can we expect this \enquote{naive} adjoining of the unitary gates to still compute the overlap of embedded states for the \ac{QEK}.

\subsection{Mitigating Depolarizing Noise} \label{sec:device_noise_mitigation}
Mitigating the effects of device noise is very important to make \ac{NISQ} practice. It is therefore no surprise that the topic has gained a lot of attention and that many techniques have been developed to mitigate device noise~\cite{temme2017error,endo2018practical,lowe2020unified,giurgica-tiron2020digital,strikis2021learning-based,koczor2021exponential,huggins2021virtual}. In the following, we will complement these with an approach that exploits the very definition of the quantum embedding kernel and that can be freely combined with other mitigation approaches.

We have introduced depolarizing noise as a rather general approach to model the noise in quantum devices.
We will model the noise with $\calD_{\lambda}$ as in Eq.~\eqref{eq:depolarizing_noise}, where the depolarizing channel is assumed to act homogeneously on the whole system.
We will refer to $\lambda$ -- the probability that the depolarizing channel does not cause a loss of information about the underlying state -- as \emph{survival probability}. Note that it may well be possible that the probabilities $\lambda_i$ differ for distinct embedded datapoints $\xx_i$, as one might need longer pulse sequences to be embedded than the other, causing more noise.

We now assume that the embedding is composed of this noise channel and the noiseless unitary embedding:
\begin{align}
    \rho(\xx) = \calD_{\lambda}[\dyad{\phi(\xx)}{\phi(\xx)}].
\end{align}
We can exploit the composition rule $\calD_{\lambda_1} \calD_{\lambda_2} = \calD_{\lambda_1 \lambda_2}$ to compute the noisy kernel matrix entries
\begin{align}
    K^{\mathrm{(dev)}}_{ij} &= \Tr \{\rho(\xx_i)\rho(\xx_j)\} \\
    &= \Tr \{\calD_{\lambda_i}[\dyad{\phi(\xx_i)}{\phi(\xx_i)}]\calD_{\lambda_j}[\dyad{\phi(\xx_j)}{\phi(\xx_j)}]\}\\
    &= \Tr \{\calD_{\lambda_i\lambda_j}[|\phi(\xx_i)\rangle\!\langle\phi(\xx_i)|\phi(\xx_j)\rangle\!\langle \phi(\xx_j)|]\} \\
    &= \lambda_i \lambda_j K_{ij} + (1-\lambda_i \lambda_j)\frac{1}{2^N}.
    \label{eq:global_noise_kernel}
\end{align}

We can exploit the fact that all diagonal entries of the noiseless kernel matrix $K$ are known to be $1$. While we could use this knowledge to save quantum computational cost, we propose to use it to gather information about the device noise. We can use Eq.~\eqref{eq:global_noise_kernel} to infer the survival probability $\lambda_i$ from the diagonal element of the noisy kernel matrix $K^{\mathrm{(dev)}}_{ii}$:
\begin{align}\label{eq:depol_rate_i}
    \lambda_i = \sqrt{\frac{K^{\mathrm{(dev)}}_{ii} - 2^{-N}}{1 - 2^{-N}}}.
\end{align}
With those values at hand we can recover the noiseless kernel matrix entries
\begin{align}
    K_{ij} = \frac{K^{\mathrm{(dev)}}_{ij} - 2^{-N}(1 - \lambda_i \lambda_j)}{\lambda_i \lambda_j}.
\end{align}

We denote this mitigation strategy as $\mitigation{SPLIT}$. We can deduce two even simpler mitigation strategies from this approach by assuming that all $\lambda_i$ have the same value. This value can be estimated by averaging multiple of the $\lambda_i$ obtained from Eq.~\eqref{eq:depol_rate_i}, a strategy which we denote as $\mitigation{MEAN}$ and which requires less diagonal elements to be measured. Alternatively, we can choose to further save resources and only measure one diagonal entry to estimate the survival probability, which we denote as $\mitigation{SINGLE}$.
There are a number of options for which entry to use, for the sake of simplicity and reproducibility we always use the first entry.

\subsection{Finite Sampling Noise}
\label{sec:sample_noise}
Recall Eq.~\eqref{eq:qek_pure_state}, where we first introduced the definition of \acp{QEK}.
And, critically, notice from Eq.~\eqref{eq:adjoint} that we proposed the so-called adjoint method for estimating the overlap: a frequentist way of approximating the kernel function from quantum circuit evaluations.
Measuring such kernel functions results in \acf{iid} Bernoulli random variables $\hat{k}_{ij}$, since each circuit evaluation outputs either a $1$, in case the observed state is $\ket{0}$, or a $0$ otherwise. By construction, the theoretical kernel value $K_{ij}$ is the true mean of this random variable, i.e., $\bbE(\hat{k}_{ij}) = K_{ij}$. 
Since it follows from Born's rule that an infinite number of circuit evaluations would be required to pin down the exact kernel value, there exists a second source of noise originating from using a finite number of samples.

In reality, we can only estimate the kernel function from a finite number of experimental runs. 
How many runs we can afford is limited by our experimental resources.
This incurs uncertainty beyond the device noise, especially if the number of runs is small.

Going one step further, notice that the pipeline involves estimating the entire kernel matrix, comprising $n(n-1)/2 = \calO(n^2)$ independent entries. To gauge the number of required circuit evaluations $M_\mathrm{tot}$ to reach a desired error $\epsilon$ in operator distance of an estimator to the target kernel matrix, we can use results from random matrix theory.

The difference between an estimator constructed using $M$ circuit evaluations \emph{per entry}, $(\bar{K}_M)_{ij} = \sum_{s=1}^M \hat{k}^{\text{(s)}}_{ij}/M$, and the target kernel matrix $K$ is given by
\begin{equation}
    (\Bar{E}_M)_{ij} = (\bar{K}_M)_{ij} - (K)_{ij} = \frac{1}{M}\sum_{s=1}^M \hat{k}^{\text{(s)}}_{ij} - K_{ij}.
\end{equation}

Remembering that
\begin{align}
    \label{eq:mean}
    \bbE\left\{(\Bar{E}_M)_{ij} \vphantom{]^4}\right\} &= 0,\\
    \label{eq:var}\bbE\left\{[(\Bar{E}_M)_{ij}]^2\right\} &= \calO\left(\frac{1}{M}\right),\\
    \label{eq:fourth_moment}\quad \bbE\left\{[(\Bar{E}_M)_{ij}]^4 \right\} &= \calO\left(\frac{1}{M^2}\right),
\end{align}
allows us to make use of the following result:
\begin{theorem}[Latala's theorem \cite{latalaEstimatesNormsRandom2005,vershyninIntroductionNonasymptoticAnalysis2011}]
Let A be a random matrix whose entries $a_{ij}$ are independent centered random variables with finite fourth moment. Then, for $C>0$,\footnote{$C$ is a constant depending only on the subgaussian norm of the entries.}
\begin{align} 
\begin{split}
    \bbE\{\Norm{A}\}\leq C\Big[&\max_i\big(\sum_j\bbE(a_{ij}^2)\big)^{1/2}\\ 
    \mathstrut+\mathstrut&\max_j\big(\sum_i\bbE(a_{ij}^2)\big)^{1/2}\\
    \mathstrut+\mathstrut&\big(\sum_{ij}\bbE(a_{ij}^4)\big)^{1/4}\Big].
\end{split}
\end{align}
\end{theorem}
For our $n\times n$-dimensional error matrix $\Bar{E}_{M}$ with moments as in Eqs.~(\ref{eq:mean}-\ref{eq:fourth_moment}), this leads to 
\begin{equation}
    \bbE\{\Norm{\Bar{E}_M}\} = \calO\left(\frac{\sqrt{n}}{\sqrt{M}}\right).
\end{equation}
Consequently, $M=\calO(n/\epsilon^2)$ measurements per kernel matrix entry are required to ensure an error of $\varepsilon$ in operator distance. As a result, since we need to estimate $\calO(n^2)$ entries of the kernel matrix, we require a total of 
\begin{equation}
    M_\mathrm{tot}=\calO\left(\frac{n^3}{\epsilon^2}\right)
\end{equation}
circuit evaluations to reach the desired accuracy.\
A short calculation for Gaussian variables using Bai-Yin's law \cite{vershyninIntroductionNonasymptoticAnalysis2011,baiLimitSmallestEigenvalue1993} verifies that this scaling is indeed asymptotically optimal, since the error of kernel matrix entries converges to the Gaussian distribution according to the central limit theorem.

The corresponding constant prefactors can be obtained via involved methods using results from random matrix theory for matrices with subgaussian rows \cite{vershyninIntroductionNonasymptoticAnalysis2011}

Another important quantity we need to estimate for quantum embedding kernels is the kernel target alignment introduced in Eq.~\eqref{eq:kernel_target_alignment_kernel_function}, which we use during the kernel training. 
Allowing for an error of $\varepsilon$ in the kernel target alignment, error propagation suggests that $\calO(1/\varepsilon^2)$ measurements per kernel entry are required, leading to $\calO(n^2/\epsilon^2)$ circuit evaluations in total.

\subsection{Regularizing the Kernel Matrix}\label{sec:regularization}
Due to the imperfect sampling outcome for the kernel matrix and the device noise mitigation techniques introduced above, the obtained matrix might not be positive semi-definite.
However, we know the exact kernel matrix to be positive semi-definite and this property is a requirement for the matrix to be used in a classification task.
We may therefore regularize the obtained matrix, validating it as kernel matrix and bringing it closer to the perfect outcome.

We discuss three methods to find a positive semi-definite matrix close to a symmetric matrix $A$:
In the first method called \emph{Tikhonov regularization}, we displace the spectrum of $A$ by its smallest eigenvalue $\sigma_{\min}$ if it is negative, by subtracting it from all eigenvalues or equivalently from the diagonal~\cite{roth2004OptimalCluster}:
\begin{align}
    \regularization{TIK}(A) = \begin{cases} 
    A - \sigma_{\min} 1 & \text{ if } \sigma_{\min}<0\\ 
    A & \text{ else } \end{cases} ,
\end{align}
which yields a positive semi-definite matrix.
While being formally the same as the original method by Tikhonov~\cite{Tikhonov1943}, we use it here to assure positive semi-definiteness instead of non-singularity of the matrix.

The second method called \emph{thresholding} only changes the negative eigenvalues of $A$ by setting them to zero~\cite{graepel1999Classification}.
This is done via a full eigenvalue decomposition, adjustment of the negative eigenvalues and composition of the adjusted spectrum and the original eigenvectors:
\begin{align}\label{eq:thresholding}
    D &= V^T A V\\
    D'_{ij} &= \max\{D_{ij}, 0\}\\
    \regularization{THR}(A) &= V D' V^T.
\end{align}
This approach is equivalent to finding the positive semi-definite matrix closest to $A$ in any unitarily invariant norm. It is also equivalent to finding the positive semi-definite matrix which has the largest alignment (see Eq.~\eqref{eq:kernel_target_alignment_kernel_matrix}) with $A$.

The third method extends this reasoning by searching the closest matrix in Frobenius norm, but with the additional requirement that the diagonal elements of the regularized matrix need be one, incorporating our knowledge about the exact kernel as a constraint. This approach constitutes an \ac{SDP} and is therefore efficiently computable:
\begin{align}
    \regularization{SDP}(A)&=\operatorname{argmin} \left\{ \lVert A'-A \rVert_F : A'\succcurlyeq 0,\; A'_{ii}=1 \right\}.
\end{align}

For further details on the computational cost and properties of the regularized matrices please refer to App.~\ref{sec:postprocessing_details}.

We note that other mitigation techniques proposed in the literature may be combined with the ones discussed here, as they function on different levels of abstraction.
An established method to reduce the impact of noise is zero noise interpolation to first order~\cite{temme2017error, kandala2018extending, havlicekSupervisedLearningQuantum2019}, which makes use of additional circuit evaluations at increased noise rates.
Furthermore, techniques to suppress errors by duplicating the circuit have been proposed recently~\cite{huggins2020virtual, koczor2020exponential}.
Both of these methods may be used to greatly reduce the noise on the kernel matrix before treating it with regularization and mitigation techniques.

During the preparation of this work, Wang \emph{et al.}~\cite{wang2021UnderstandingQEKPower} demonstrated that regularization methods can improve the classification accuracy of noisy circuits significantly, more concretely $\regularization{TIK}$, $\regularization{THR}$ and flipping the negative eigenvalues of the kernel matrix were covered.

\section{QEK Pipeline}\label{sec:qek_pipeline}

\begin{figure*}
    \centering
    \includegraphics{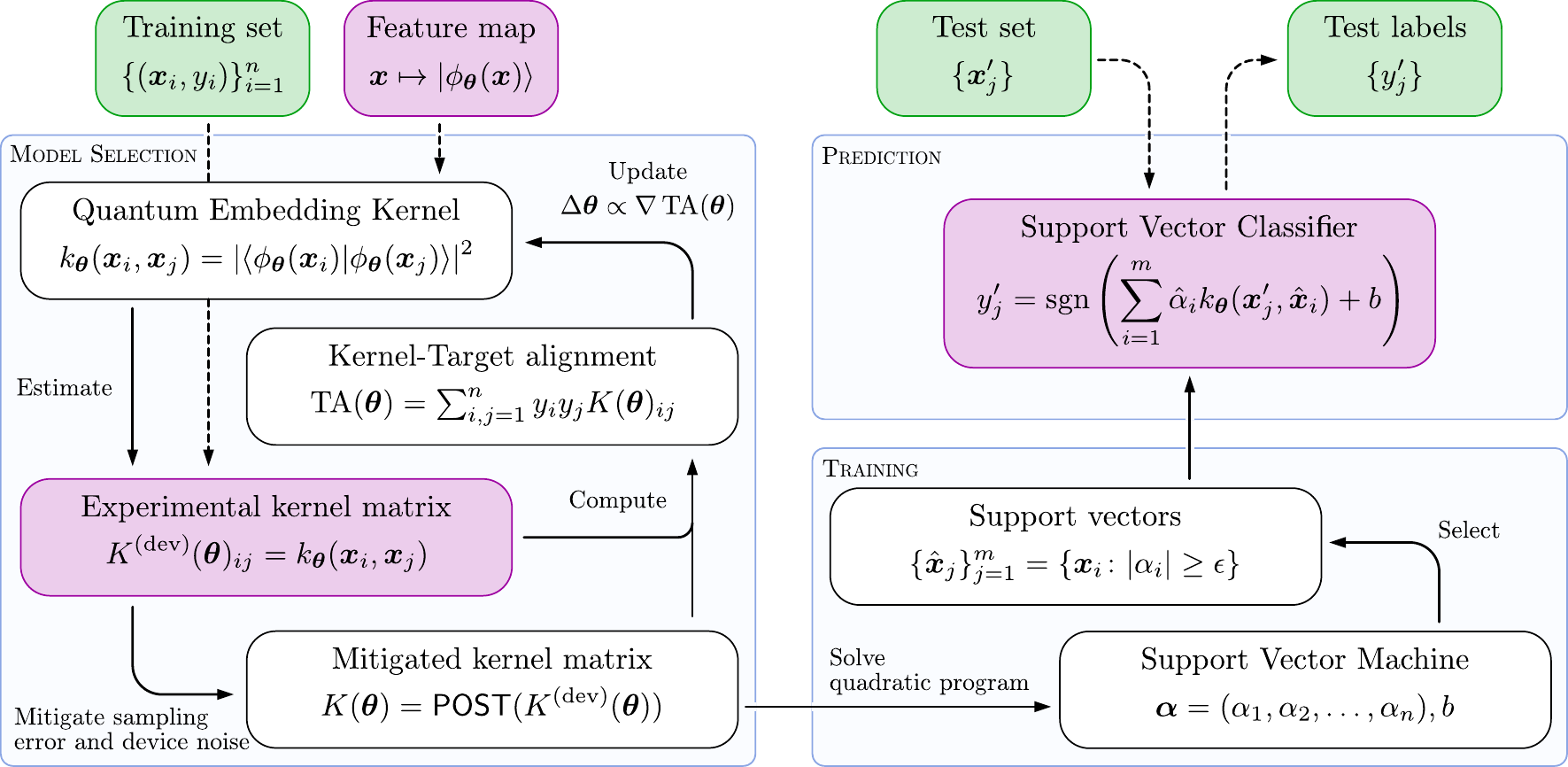}
    \caption{Schematic of the pipeline used in this work. Green boxes indicate data, purple boxes indicate process steps that are executed on quantum hardware. The pipeline used in this work can be split into three separate parts. In the model selection part, depicted on the left, the parameters of the feature map are adjusted to increase the kernel-target alignment. To calculate the alignment, the kernel matrix is computed and may afterwards be post-processed to mitigate sampling and device noise. After a sufficient target-alignment is reached, the kernel is used to train a support vector machine. The resulting support vector classifier is used in the prediction step to predict labels of new datapoints.}
    \label{fig:pipeline}
\end{figure*}

The approaches presented in the previous sections can be combined into a holistic pipeline for working with quantum embedding kernels as depicted in Fig.~\ref{fig:pipeline}. We start from a parameterized quantum circuit that represents a quantum feature map with variational parameters set to some initial values. To adjust the parameters for a specific dataset, a training loop is run: First, a kernel matrix is obtained from the underlying NISQ device. As an alternative next step, a mitigation and regularization strategy can be applied to improve the quality of the kernel matrix. The kernel matrix is then used to calculate the kernel-target alignment and its gradient with respect to the variational parameters of the parameterized quantum circuit. Gradient descent is then used to update the variational parameters and hence the quantum embedding kernel. This process is repeated until the desired kernel-target alignment is reached.

To perform a classification of new data, a support vector machine is trained using the post-processed kernel matrix of the optimized quantum embedding kernel. The support vectors of the SVM are then extracted and can be used in a support vector classifier to predict labels for new datapoints. To this end, the quantum embedding kernel between the new datapoints and the support vectors have to be computed, but the training of the SVM itself is purely classical.

\section{Numerical Experiments}
\label{sec:numerical_experiments}
Let us make our discussion concrete by designing and performing some proof-of-principle experiments. We consider both noiseless and noisy simulations of quantum circuits as well as experiments on actual \ac{NISQ} devices.
Each of our experiments is designed to illustrate a specific step of the main pipeline depicted in Fig.~\ref{fig:pipeline}.
Accordingly, we use different datasets and \acp{QEK} curated to each experiment and repetition.

For the sake of illustration we choose to work on datasets of $2$ dimensions. 
We use two artificial and one semi-artificial dataset.
The exact construction of each dataset is detailed in App.~\ref{app:numerical_experiments}.

The \emph{checkerboard} dataset represents a $4\times4$ grid of alternating classes, where the elements of the checkerboard are drawn from a continuous uniform distribution centered in the tiles of the checkerboard. 
Due to its many different connected components, the checkerboard dataset requires a medium sized \ac{QEK}, which we use for benchmarking our mitigation and regularization techniques.
Next, we sample pixels from MNIST handwritten digit images of classes $0$ and $1$ and generate datasets \emph{zero vs. not-zero}, \emph{one vs. not-one}, \emph{base-zero} and \emph{base-one} to show a more realistic use of \acp{QEK}, for which we use much larger embedding circuits.
Finally, we have the \emph{symmetric donuts} dataset, consisting of two pairs of circumscribed circles with alternating classes.
We use real hardware for this dataset and thus use a fairly shallow and narrow \ac{QEK}. 

For our experiments, we always used the adjoint approach because of its lower qubit requirements. 
For the quantum feature map, we follow Refs.~\cite{perez-salinasDataReuploadingUniversal2020, schuldEffectDataEncoding2020, mitarai2018CircuitLearning} and use a \emph{data re-uploading} quantum embedding circuit. 

The circuit used in this work is a repetition of the following layer architecture that is responsible for both embedding data and applying trainable gates.
One layer comprises: one layer of Hadamard gates; one layer of single-qubit Pauli-$Z$ rotations, embedding one data feature each; one layer of trainable single-qubit Pauli-$Y$ rotations; and one \emph{ring} of controlled Pauli-$Z$ rotations, in which each qubit acts as control for the rotation on the next adjacent qubit, considering first and last as adjacent. Note that all controlled Pauli-$Z$ rotations mutually commute, therefore allow to execute the ring in constant circuit depth.
Consequently, for a circuit constructed using $L$ blocks and $N$ qubits, there are $2NL$ trainable parameters.

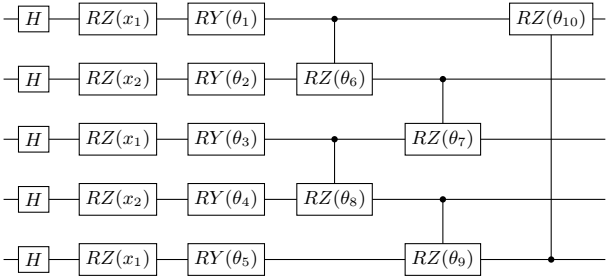
\begin{figure}
    \input{figures/QEK_block_tikz}
    \caption{Circuit diagram of the elementary building block used in the \ac{QEK} ansatz for $N=5$ qubits and $m=2$ features. The trainable parameters of the ansatz are denotes as $\{\theta_j\}_{j=1}^{s=10}$ and $x_{1,2}$ are the data features.}
    \label{fig:circuit_layer}
\end{figure}

Furthermore, we do not tie the $i^\text{th}$ feature to one particular qubit, but rather iterate over the features cyclically in each embedding layer.   
That means, if there are more data features $n$ than qubits in the circuit $N$, one block is not enough to encode all features.
In this case, we encode the first $x_1,\ldots,x_N$ features in the first block, and for the second block we start encoding features $x_{N+1}, x_{N+2}$, and so on until we reach $x_n$.
From there on, provided there are still more encoding gates available in the circuit, we start over from $x_1$.
The circuit diagram for a single block encoding $2$-dimensional data on $5$ qubits is shown in Fig.~\ref{fig:circuit_layer}.

When training our models, we treat the number of qubits $N$ and the number of blocks $L$ as hyperparameters of the problem, much like the number of neurons per layer and the number of layers in artificial neural networks. In \ac{QML}, the number of qubits is usually referred to as \emph{width}, and the number of blocks as \emph{depth}.

With these kernels, we perform three experiments, corresponding to the three sections that follow.
In Sec.~\ref{ssec:noiseless_simulations} we check the validity of training \acs{QEK} via noiseless simulations.
Next, we allow for noise sources (as considered in Sec.~\ref{sec:the_effects_of_noise}) to test the aforementioned mitigation and regularization techniques in Sec.~\ref{ssec:numerics_mitigation}, still on a classical simulator. 
Finally, the culminating experiment studies mitigation techniques again, but this time using IonQ's \ac{QPU}~\cite{wright2019ionq}, to demonstrate functionality on an actual quantum device.

Our experiments were implemented using the PennyLane library for \ac{QML}~\cite{bergholm2018pennylane} and code is available at~\cite{hubregtsen2021code}. Hardware experiments were executed on an IonQ device~\cite{wright2019ionq} via Amazon Braket~\cite{aws2021braket}.
    
\subsection{Noiseless simulations} \label{ssec:noiseless_simulations}
We kick off by testing our approach on ideal lab conditions.
As a first step, we assume both completely negligible noise and access to infinite samples.
This can be achieved using a classical simulator that stores the entire wave vector at every step.
We perform several experiment repetitions using all introduced datasets.

The experiment takes three inputs: the hyperparameter values for the \ac{QEK}, the training set, and the test set.
Once these are provided, the steps are always:
(1) sample a few sets of kernel parameters at random (we do $3$ or $5$).
We refer to the kernels using these sets of parameters as \emph{untrained} kernels.
(2) to each untrained kernel, fit a \ac{SVM} using the training set.
Denote the resulting classification accuracy on test as \emph{untrained} accuracy.
(3) select one of the untrained kernels based on some quality of its respective accuracy (we select the lowest one).
(4) tune the parameters of the selected untrained kernel by maximizing the kernel target alignment (we use for instance Stochastic Gradient Ascent, with a batch size as small as $4$).
Once optimization is finished, we call the resulting object the \emph{trained} kernel.
(5) fit one last \ac{SVM} using the trained kernel and the training set and report the achieved accuracy on the test data.
This is the \emph{trained} accuracy.
Of our interest in this experiment are, as said, the untrained and trained accuracy.

There is one additional experiment designed specifically for the semi-artificial MNIST dataset.
For semi-artificial MNIST, we train the kernels only on the zero vs. non-zero and one vs. non-one datasets.
We tackle a \enquote{Zero vs. one} membership problem with the base-zero and base-one sets via a majority vote ensemble of the trained kernels from last step.
More extensive explanation on the ensemble experiment to be found in App.~\ref{app:numerical_experiments}.

According to the theory presented in Sec.~\ref{sec:training_quantum_embedding_kernels} and in~\cite{wangOverviewKernelAlignment2015}, one should expect that a trained \ac{QEK} should outperform the randomly initialized kernel pre-training\footnote{
Important noting is that target alignment optimization is a heuristic that aims at guaranteeing \emph{good} classification and generalization, not \emph{better}.
There exist adversarial datasets where perfect accuracy can be achieved with very low alignment.
For instance, consider a dataset where points of each class are organized in two long parallel lines, one label per line.}.
We expect that larger \ac{QEK} profit more from training, because deeper and wider systems have a larger parameter space, and random initialization therefore has a lower chance of falling close to a local maximum.
More trainable parameters also convey more expressivity.

\begin{figure}
    \centering
    \includegraphics{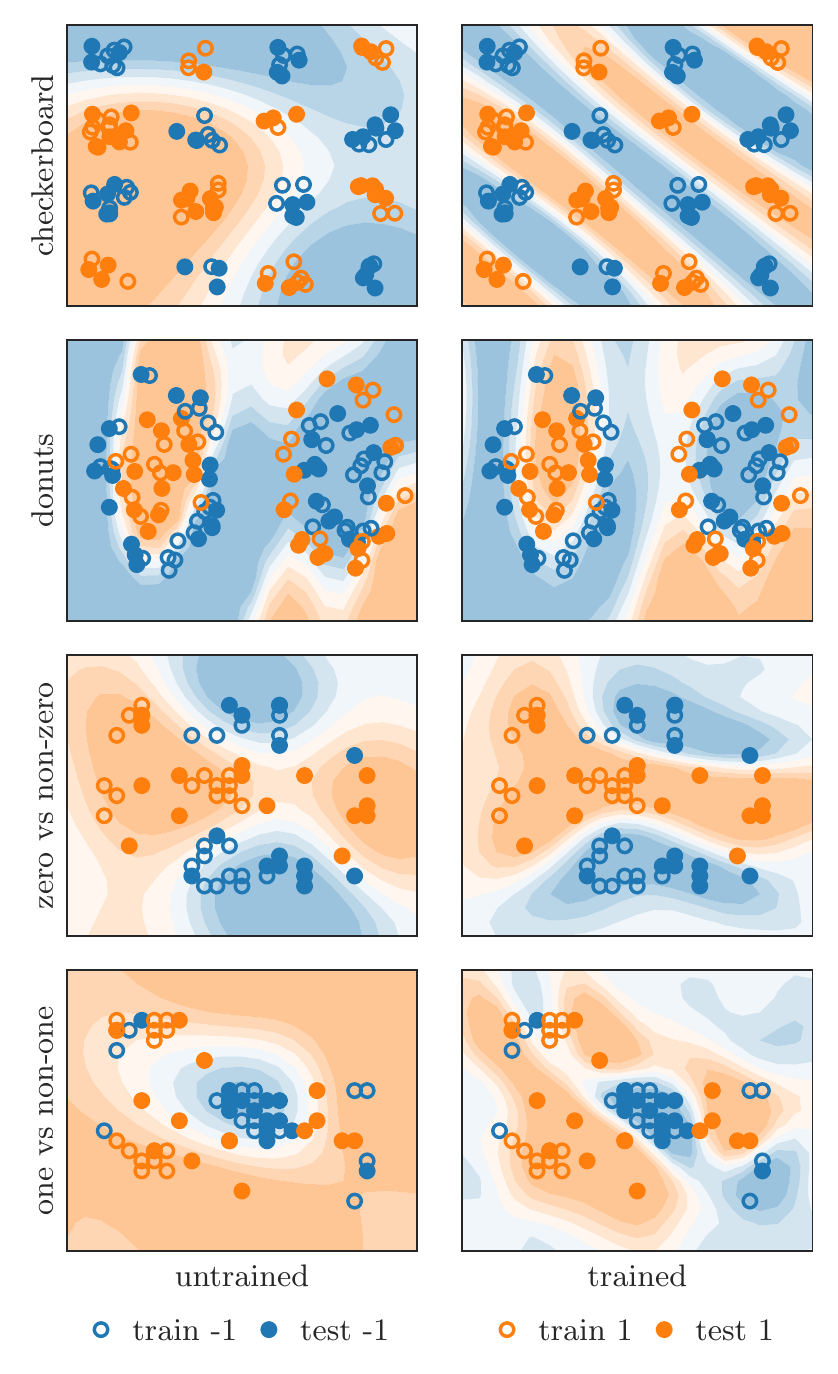}
    \caption{
    Datasets and decision boundaries of \acp{SVM} constructed using different \acp{QEK} in noiseless simulations.
    Each plot shows the real plane and the axes are cartesian coordinates in arbitrary units.
    The rows correspond to the different datasets, of which a detailed account can be found in App.~\ref{app:numerical_experiments}.
    The columns correspond to the kernel before (left) and after (right) target alignment training.
    The classification accuracy achieved in every repetition is presented in Tab.~\ref{tab:kernel_experiments}, next to the respective hyperparameters.}
\label{fig:datasets_and_decision_boundaries}
\end{figure}

Fig.~\ref{fig:datasets_and_decision_boundaries} shows how the quality of the decision boundaries improves from the untrained kernels (left column) to the trained ones (right column).
In all experimental repetitions, though, there are still some misclassified points, which is not unexpected as the underlying kernel functions likely have limited expressivity.
The precise numerical results of this experiment are found in Tab.~\ref{tab:kernel_experiments}.
The table lists the minimum and maximum untrained accuracy with random parameters for the kernel, the trained accuracy and the choice of hyperparameters for each experiment repetition. 
Notice that the \enquote{zero vs. one} ensemble experiment does not have untrained kernels by construction.

Upon inspecting the results, we notice how trained accuracy values range from $0.75$ for the smallest \ac{QEK} (width and depth equal to $3$) to $0.97$ for repetitions with larger circuits and we even obtain a perfect score for the ensemble experiment.
We identify the anticipated general trends: training the kernel always improved the accuracy compared to the untrained kernel with minimum accuracy in our experiment.
This is empirical evidence that maximizing kernel target alignment comes with improved accuracy.
Despite the consistent improvement over the untrained kernel with minimum accuracy, the untrained kernel with maximum accuracy matches the trained kernel in the ``zero vs. not-zero" repetition, and even comes out on top for the smaller instance of ``symmetric donuts".
Note that increasing the size (and implicitly the expressivity) of the embedding circuit in the ``symmetric donuts" repetition leads to the \emph{trained} \ac{QEK} accuracy being higher than the maximum \emph{untrained} one.

\begin{table*}
\caption{Kernel experiments and ensemble experiments - datasets, circuit hyperparameters and achieved accuracies} \label{tab:kernel_experiments}
\begin{tabular*}{\textwidth}{c @{\extracolsep{\fill}} ccccccc }
\toprule
 Dataset & total samples ($n$) & Width ($N$) & Depth ($L$) & Untrained & Untrained & Trained \\
  & & & & (minimum) & (maximum) & (from minimum) \\ 
  \midrule
 Checkerboard & $60$ datapoints & $5$ & $8$ & $0.52$ & $0.52$ & $0.97$\\  
 Zero vs not-zero & $30$ datapoints & $4$ & $32$ & $0.9$ & $0.97$ & $0.97$ \\
 One vs not-one & $30$ datapoints & $4$ & $32$ & $0.77$ & $0.77$ & $0.97$ \\
 Zero vs one & $10$ images ($15$ datapoints each) & $4$ & $32$ & NA & NA & $1$\\
 Symmetric Donuts & $120$ datapoints & $3$ & $3$ & $0.58$ & $0.82$ & $0.75$  \\
 Symmetric Donuts & $120$ datapoints & $4$ & $3$ & $0.65$ & $0.70$ & $0.85$\\
 \bottomrule
\end{tabular*}
\end{table*}

\subsection{Mitigation and regularization experiments}\label{ssec:numerics_mitigation}

\begin{figure}
    \centering
    \includegraphics[width=0.48\textwidth]{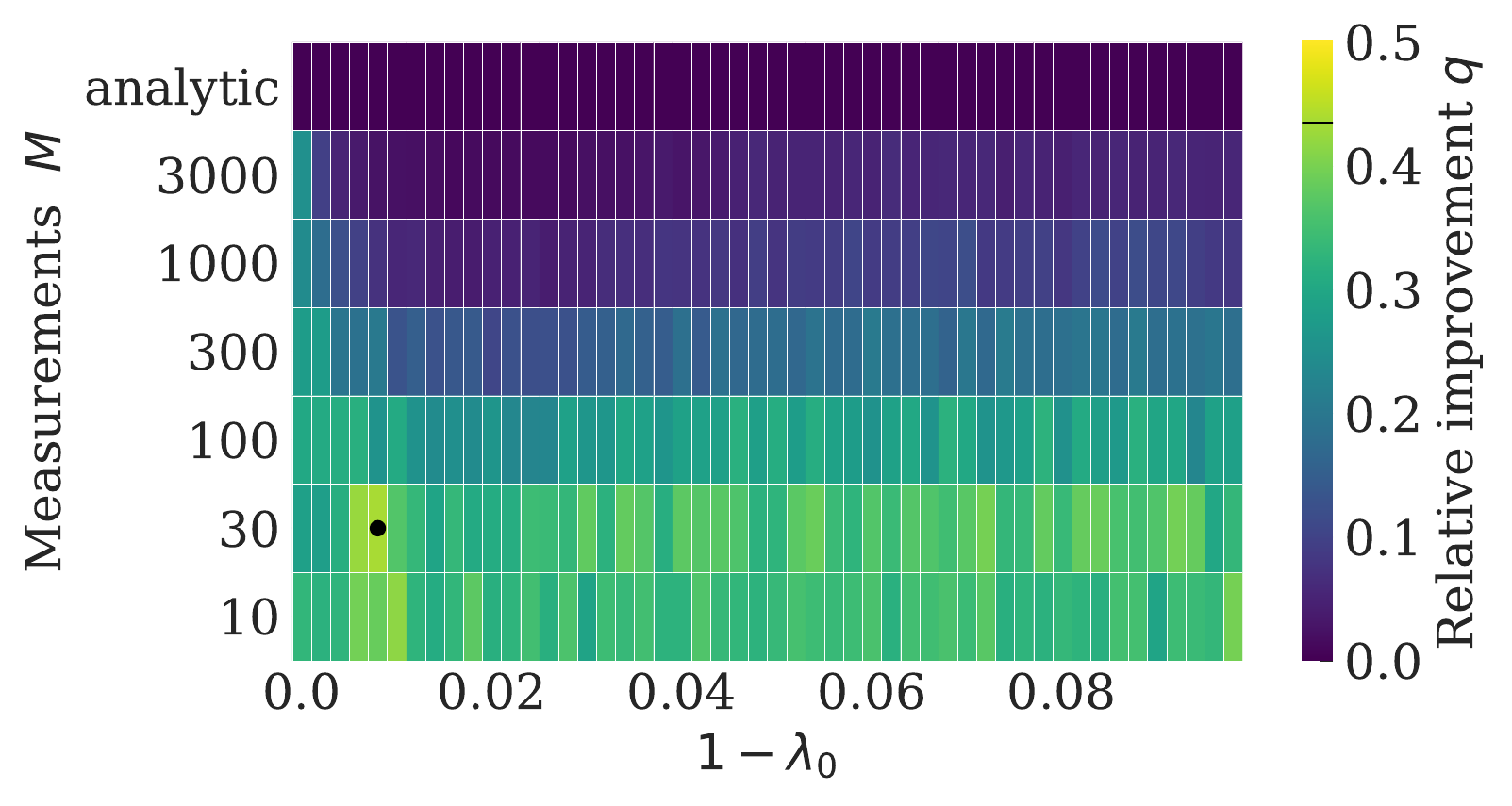}
    \caption{Relative improvement $q$ in alignment (see Eq.~\eqref{eq:def_quality_postproc}) for several base survival probability $\lambda_0$ and numbers of circuit evaluations per matrix entry $M$ under a noise model based on local depolarizing noise (see App.~\ref{sec:noise_model}).
    We here use the best of our post-processing combinations, which consists only of $\regularization{SDP}$, rated by how often they performed best across all noise regimes (for the detailed ranking see App.~\ref{sec:postprocessing_rating}).
    The additional line in the color bar and the marker in the heatmap indicate the best relative improvement and the corresponding noise parameters.
    }
    \label{fig:relative_improve_postprocessing}
\end{figure}

We now investigate the effect of the regularization and device noise mitigation techniques introduced in Sec.~\ref{sec:device_noise_mitigation} and~\ref{sec:regularization}.
To this end, we simulate device noise with a noise model based on local depolarizing noise (see App.~\ref{sec:noise_model} for details) and test the post-processing performance on the checkerboard dataset.

For a range of base survival probabilities $\lambda_0$ and measurements $M$ per kernel matrix entry, we first compute the noisy, sampled kernel matrix $\Bar{K}_M$.
We then consider any combination of up to three methods with the order \emph{regularize - mitigate - regularize}, including combinations that skip one or two of these steps, to post-process $\Bar{K}_M$ into $K^\mathrm{(post)}$.
The quality of each mitigation strategy is finally determined as the change in alignment with $K$ (introduced in Eq.~\eqref{eq:matrix_alignment}), the noiseless matrix, compared to $\Bar{K}_M$, relative to the deviation of the raw matrix from perfect alignment:
\begin{align}\label{eq:def_quality_postproc}
    q = \frac{\A\!\left(K^\mathrm{(post)}\right)-\A\!\left(\Bar{K}_M\right)}{1-\A\!\left(\Bar{K}_M\right)},
\end{align}
where in a slight abuse of notation we abbreviated $\A\!\left(K^\mathrm{(post)}\right)\coloneqq \A\!\left(K^\mathrm{(post)}, K\right)$ and skipped the dependence on the exact kernel matrix $K$.

Fig.~\ref{fig:relative_improve_postprocessing} shows the improvement $q$ across the base survival probabilities and numbers of circuit evaluations for the best of the $42$ combinations of mitigation and regularization strategies, achieving values between $0\%$ and $43.5\%$. We especially see larger improvements for lower numbers of measurements.
For more details on the considered combinations of regularization and mitigation and for the best choice for different noise regimes, please refer to App.~\ref{sec:postprocessing_rating}.

Overall, we see that post-processing can in general enhance the quality of the obtained kernel matrix significantly, in addition to other possible mitigation techniques that may reduce the sampling and device noise strengths effectively~\cite{temme2017error, kandala2018extending, havlicekSupervisedLearningQuantum2019} or even at the hardware level~\cite{koczor2020exponential, huggins2020virtual}.
At the same time, one has to choose the mitigation and regularization techniques carefully as their quality depends on the number of circuit evaluations and device noise level.
In particular, the post-processing methods are not attributed to one source of noise each, but e.g.\ regularization might be the better strategy for tackling device noise, which it was not designed to counter, as we saw in the regime of strong device noise.
If no estimate for the noise levels is available, applying a simple regularization routine such as thresholding offers systematic and consistent improvement, but for realistic noise levels, device noise mitigation typically provides even better results (also see Sec.~\ref{ssec:numerics_hardware} and App.~\ref{sec:postprocessing_app}).

When using one of the mitigation techniques, one should consider that for $\mitigation{SINGLE}$, only one of the diagonal entries of the kernel matrix needs to be calculated, while multiple (all) diagonal entries are required for $\mitigation{MEAN}$ ($\mitigation{SPLIT}$).

\subsection{Hardware experiments}\label{ssec:numerics_hardware}
So far, all experiments in this section were run on classical simulators.
While we tried to use fair noise models and reasonable circuits, it is worthwhile to investigate the behavior in real-world conditions.
To test the performance of the introduced techniques on actual quantum hardware, we have computed the kernel matrix for the symmetric donuts dataset using three qubits on an ion trap \ac{QPU} by IonQ.

For the computation we have used $M=175$ circuit evaluations per kernel matrix entry and because we measured the diagonal entries for mitigation purposes, the total number of circuit evaluations sums up to about $3.2\cdot 10^{5}$ in total. 
In addition, we have sampled kernel matrices for several smaller $M$ values from the measured distribution\footnote{Note that this is not the same as a proper computation on the quantum device with decreased $M$ because we sample from a sample and not from the true distribution directly.}.

\begin{figure}
    \centering
    \includegraphics[width=0.48\textwidth]{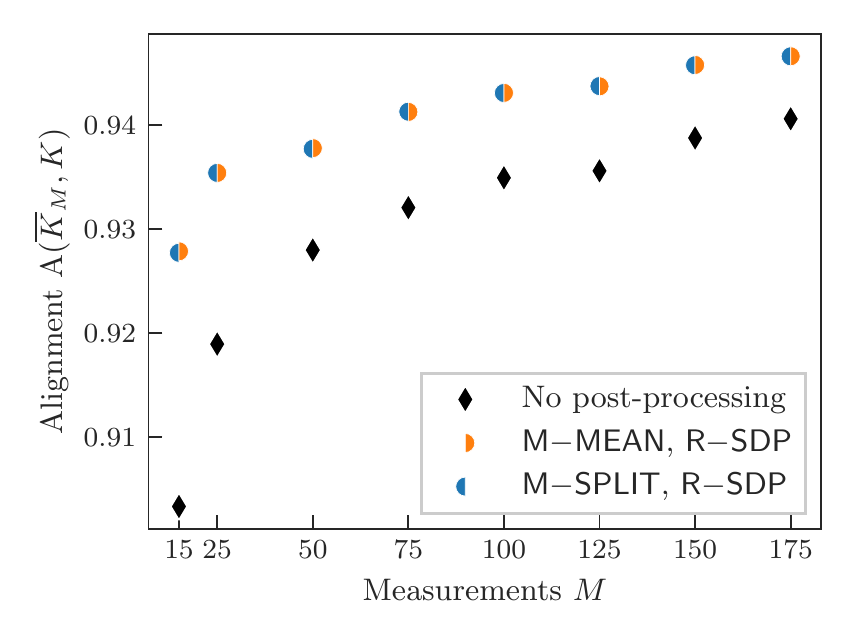}
    \caption{Alignment $\A$ of the kernel matrix measured on the ion trap \ac{QPU} with the simulated, noiseless kernel matrix $K$ for various numbers of circuit evaluations per matrix entry $M$, with and without the respective two best post-processing strategies.
    Applying our device noise mitigation techniques $\mitigation{MEAN}/\mitigation{SPLIT}$ (see Sec.~\ref{sec:device_noise_mitigation}), which assume a simple, global depolarizing noise model, and matrix regularization $\regularization{SDP}$ results in an improved alignment.}
    \label{fig:ionq_mitigation}
\end{figure}

Fig.~\ref{fig:ionq_mitigation} shows the alignment $\A\!\left(\Bar{K}_M\right)$ between the obtained kernel matrix $\Bar{K}_M$ and the noiseless matrix $K$, as well as the alignment $\A\!\left(K^\mathrm{(post)}\right)$ between the post-processed matrix $K^\mathrm{(post)}$ and $K$.
For each number of circuit evaluations $M$, we plot the two best of the $42$ post-processing combinations. Note that many of these combinations yield a quality similar to the best choice.
As expected, the quality of the kernel matrix improves with the number of circuit evaluations and as predicted by our simulation results (see App.~\ref{ssec:numerics_mitigation}), the post-processing methods increase the alignment significantly.
The achieved values for the relative improvement $q$ range between $10.1\%$ and $25.4\%$ with a mean of $14.9\%$.

We note that the combination $\mitigation{MEAN},\;\regularization{SDP}$, which is either best or second best in the hardware results, was correctly predicted for small device noise levels by our simulations of depolarizing noise, on a different dataset, with different circuit depth and width (see App.~\ref{sec:postprocessing_app}) and compiled to a different elementary gate set.
This indicates that the depolarizing noise model captures properties of the noise in the \ac{QPU} that are significant for the kernel matrix computation, and suggests that these post-processing methods show robust performance across different circuit depths, qubit numbers and datasets.

In conclusion, our results on the actual quantum device demonstrate an increased kernel matrix quality when using post-processing, which may allow for improved classification accuracy (also see~\cite{wang2021UnderstandingQEKPower}) or alternatively for a reduced number of circuit evaluations while maintaining a fixed classification performance.

\section{Summary and Outlook}
\label{sec:summary_and_outlook}
\acresetall

In this work, we have studied the concept of \acf{QEK}. 
Here, quantum embedding circuits serve as feature maps for kernel-based \ac{ML}.
To optimize variational parameters of the \acp{QEK}, we transferred the concept of kernel target alignment to the quantum setting. We have performed various numerical experiments that showed improvement in classification accuracy after training.

As the kernel matrices must be positive semi-definite, and the \acp{QEK} run on near-term quantum device, we have also investigated noise mitigation techniques. 
Concretely, we proposed device noise mitigation techniques specific for kernel matrices and combined them with regularization methods.
We tested a large set of combinations, both on simulated depolarizing noise as well as on data from a real \acl{QPU}.
In both scenarios we found that post-processing methods can partly recover the noiseless kernel matrix.
Based on these results we recommend best-practice post-processing strategies for different noise regimes.

There are two immediate challenges remaining when applying post-processing methods. On the one hand, we need to rate the methods when we do not have access to the noiseless matrix. On the other hand, the impact of the methods on the classification accuracy remains to be investigated.

In the design and training of \acp{QEK}, one could explore various aspects. 
A clear question would be the choice of ansatz families.
Some key objects of study for this would be the expressivity of different circuits, the dependence on the dataset, the optimal choice of hyperparameters (or, alternatively, how one could perform empirical risk minimization successfully), or how one would build gauge invariant kernel functions~\cite{Bronstein2021GeometricDeepLearning}.
Another major topic is investigating the effect of the barren plateau phenomenon~\cite{mccleanBarrenPlateausQuantum2018a,cerezo2021cost,uvarov2020barren} in the kernel setting, and subsequently the study of (quantum-aware) cost function alternatives to the target alignment.

Finally, one could explore whether the proposed model can be transferred to more general tasks such as unbalanced binary classification, multi-class classification, or regression.

\section*{Acknowledgments}
The authors wish to thank Xanadu for organizing QHack 2021, where the foundations of this work were laid as part of the Open Hackathon Challenge, and the resulting funding. We further want to thank the AWS team for their support and funding that provided us access to the Rigetti and IonQ devices, as well as Sandbox@Alphabet for alpha access to the Floq cloud service, yielding access to the TPU-based quantum simulator.
Additionally, we would like to thank Richard Kueng for valuable input on bounds, as well as Jens Eisert and Maria Schuld for valuable feedback.
We endorse Scientific CO$_2$nduct~\cite{conduct} and provide a CO$_2$ emission table in App.~\ref{app:co2}. 

This work was supported by the BMWi under the PlanQK initiative, the BMBF under the RealistiQ initiative, the Cluster of Excellence MATH+ project EF1-7, the European Flagship project PasQuanS and the DFG under Germany’s Excellence Strategy Cluster of
Excellence Matter and Light for Quantum Computing (ML4Q) EXC2004/1 390534769 and the CRC 183 project B01.

\section*{Author Contributions}
JJM, EGF and PKF built the theory on trainable QEKs.
TH and EGF ran the numerics for the trainable QEKs.
DW, PKF and JJM built the theory on noise mitigation.
DW and PJHSD ran the numerics for noise mitigation.
JJM supervised the project.
All authors contributed to the discussions and to writing the manuscript. 
\bibliography{main.bib}

\appendix

\section{Connection to quantum feature map optimization}\label{sec:app_feature_map_opt}
Optimizing kernels using the kernel-target alignment as a cost function is closely related to the \enquote{metric learning} approach put forward for the training of quantum feature embeddings in Ref.~\cite{lloyd2020QuantumEmbeddingsML}. 

To understand this approach, we first introduce some notation. We consider a dataset $\calS = \{ (\xx_i, y_i) \}$ that we split in two parts corresponding to the two classes labeled as $\pm 1$. We denote these subsets as $\calS_{+}$ and $\calS_{-}$, respectively. For a given embedding $\ket{\phi_{\ttheta}(\xx)}$, we can identify both classes with quantum states -- we will refer to them as \emph{class states} -- simply by averaging the embedded quantum states
\begin{align}
    \rho_{\pm}(\ttheta) &= \frac{1}{|\calS_{\pm}|} \sum_{\xx \in \calS_{\pm}} | \phi_{\ttheta}(\xx) \rangle \! \langle \phi_{\ttheta}(\xx) | \\
    &= \frac{1}{|\calS_{\pm}|} \sum_{\xx \in \calS_{\pm}} \phi_{\ttheta}(\xx).
\end{align}
Here, we denoted the density matrix of the embedding as $\phi_{\ttheta}(\xx) = | \phi_{\ttheta}(\xx) \rangle \! \langle \phi_{\ttheta}(\xx) |$. The state $\rho_{\pm}$ models an approach where the the encoded datapoint $\xx$ is uniformly sampled from $\calS_{\pm}$.

Ref.~\cite{lloyd2020QuantumEmbeddingsML} suggests to optimize the embedding $\ket{\phi_{\ttheta}(\xx)}$ by maximizing the Hilbert-Schmidt distance of the class states, i.e.,
\begin{align}\label{eq:hs_distance}
    P(\ttheta) = \Tr\{(\rho_{+}(\ttheta) - \rho_{-}(\ttheta))^2 \}.
\end{align}
Its relation to kernel-target alignment becomes apparent if we rewrite the numerator of the kernel-target alignment -- the polarity -- in terms of these density matrices. We therefore consider the polarity for imbalanced datasets, where we rescale the labels with the number of datapoints in the class. The rescaled labels are denoted as $\hat{y}_j$.
\begin{align}
    \sum_{i,j=1}^N \hat{y}_i \hat{y}_j  k_{\ttheta}(\xx_i, \xx_j)
    &= \sum_{i,j=1}^N \hat{y}_i \hat{y}_j \langle \phi_{\ttheta}(\xx_i), \phi_{\ttheta}(\xx_j) \rangle \\
    &= \left\langle \sum_{i=1} \hat{y}_i \phi_{\ttheta}(\xx_i), \sum_{i=1} \hat{y}_i \phi_{\ttheta}(\xx_i) \right\rangle \\
    &= \left\lVert \sum_{i=1} \hat{y}_i \phi_{\ttheta}(\xx_i) \right\rVert^2.
\end{align}
The polarity is therefore nothing else but the squared norm of $\sum_{i=1} \hat{y}_i \phi_{\ttheta}(\xx_i)$, which is a weighted sum of the embedded datapoints. For \acp{QEK}, this is equal to the difference of the two class matrices introduced above:
\begin{align}
    \sum_{i=1} \hat{y}_i \phi_{\ttheta}(\xx_i) &= \sum_{\xx_{+} \in \calS_{+}} \frac{\phi_{\ttheta}(\xx_{+})}{|\calS_{+}|} - \sum_{\xx_{-} \in \calS_{-}} \frac{\phi_{\ttheta}(\xx_{-})}{|\calS_{-}|}\\
    &=\rho_{+}(\ttheta) - \rho_{-}(\ttheta).
\end{align}
This means that the polarity is equal to the Hilbert-Schmidt distance introduced in Ref.~\cite{lloyd2020QuantumEmbeddingsML}, as found in Eq.~\eqref{eq:hs_distance}

As already noted in Ref.~\cite{lloyd2020QuantumEmbeddingsML}, the polarity can be rewritten as
\begin{align}
    P(\ttheta) = \Tr\{\rho_{+}(\ttheta)^2 + \rho_{-}(\ttheta)^2 - 2 \rho_{+}(\ttheta) \rho_{-}(\ttheta) \}.
\end{align}
Consequently, increasing the polarity translates to an increase in the \emph{purity} of the class states $\Tr \{ \rho_{\pm}(\ttheta)\}^2$, thereby encouraging points in the dataset to cluster closer together in feature space. At the same time, this cost function decreases the overlap of the two data embedding states, thereby encouraging them to reside in different corners of the Hilbert space.

However, we are of the opinion that the kernel-target alignment -- representing the normalized polarity -- is a measure that is easier to interpret and more accessible to numerical optimization than the pure polarity. Ref.~\cite{lloyd2020QuantumEmbeddingsML} proposes a classifier where the overlap of the embedded datapoint with the two class states is computed. The label of the class state with the larger overlap is then assigned to the new datapoint. This corresponds to a kernelized nearest-centroid classification. We conclude that the use of the embedding in a support vector machine allows for more sophisticated decision boundaries than the method proposed in Ref.~\cite{lloyd2020QuantumEmbeddingsML}.

\section{Details on post-processing methods}\label{sec:postprocessing_app}
\subsection{Runtimes and output properties}\label{sec:postprocessing_details}
The post-processing methods we introduced in Sec.~\ref{sec:regularization} and~\ref{sec:device_noise_mitigation} differ in their classical and quantum computational cost and in the properties of the output matrix.

The regularization methods $\regularization{TIK}$ and $\regularization{THR}$ require the computation of the smallest eigenvalue and of the full eigenvalue decomposition respectively, which has classical complexity $\mathcal{O}(n^3)$ with naive methods but more realistically scales like matrix multiplication for relevant sizes with $\mathcal{O}(n^{2.8})$ (Strassen algorithm~\cite{strassen1969gaussian})\footnote{If this was to be a bottle neck, the full matrix multiplication may be skipped when multiplying the kernel matrix with vectors only.}.
The worst case scaling for $\regularization{SDP}$ is $\mathcal{O}(n^{3.8})$, again assuming the Strassen algorithm for matrix multiplication and considering that we use $n$ constraints to fix the diagonal entries~\cite{lee2015faster, strassen1969gaussian}.
In our experiments on datasets with $60$ datapoints, the former two methods had negligible computational cost, whereas $\regularization{SDP}$ took $0.5s$ on average for this rather small matrix.
In addition to this large difference in the prefactor, some additional tests for random matrices confirmed a significantly worse scaling of the cost for $\regularization{SDP}$ compared to $\regularization{TIK}$ and $\regularization{THR}$.

As they only act on the kernels spectrum, $\regularization{TIK}$ and $\regularization{THR}$ preserve the eigenbasis of the kernel matrix, a potentially relevant property for the classification task.
On the contrary, $\regularization{SDP}$ does not preserve the eigenbasis but ensures that the output kernel matrix has the correct diagonal entries.

For the proposed mitigation methods, additional quantum computation is required in order to determine the diagonal entries, which in turn are used to estimate the depolarizing survival probabilities.
The number of required entries is $1$, $n_\mathrm{mean}\in[1,n]$ and $n$ for $\mitigation{SINGLE}$, $\mitigation{MEAN}$ and $\mitigation{SPLIT}$, respectively, which then should be measured as often as the other matrix entries.
While estimating the probabilities has negligible cost, the modification of the matrix requires $\mathcal{O}(n^2)$ classical computation resources\footnote{This may again be improved if we are not interested in the fully computed matrix but e.g.\ in multiplying it with vectors, should it ever become a major resource requirement.}.

Considering Eq.~\eqref{eq:depol_rate_i}, we see that our mitigation methods estimate the survival probability $\lambda_i$ to be larger than one for $K^\mathrm{(dev)}_{ii}>1$ and to be imaginary if $K^\mathrm{(dev)}_{ii}<2^{-N}$, both being unreasonable estimates.
The first will only ever occur if a previous post-processing method increased the diagonal element $K^\mathrm{(dev)}_{ii}$ too far, as a \ac{QPU} itself will not output measurement probabilities above $1$.
The second may occur in the presence of very strong noise that suppresses the exact value of $1$ to $2^{-N}$, which would presumably imply the \ac{QPU} output to be impracticably flawed anyways. 
For $\mitigation{SINGLE}$ ($\mitigation{MEAN}$), the same reasoning holds for the single measured entry (for the average of the considered diagonal entries), i.e.~in particular for $\mitigation{MEAN}$ we are unlikely to run into either of the above problems.

Even if the estimated survival probabilities $\lambda_i$ lie in the physically meaningful range $[0,1]$, the mitigation might still produce kernel matrix entries that are not valid probabilities and thus can impossibly be the result of a real \ac{QEK} evaluation.
For a given noisy matrix entry $K^\mathrm{(dev)}_{ij}$, this happens if 
\begin{align}
    K^\mathrm{(dev)}_{ij} \not\in \left[ \varepsilon, \lambda_i\lambda_j+\varepsilon \right],
\end{align}
where we abbreviated $\varepsilon=2^{-N}(1-\lambda_i\lambda_j)$ and the estimated probabilities fulfill $\lambda_{i}=\lambda_j$ for $\mitigation{SINGLE}$ and $\mitigation{MEAN}$.
Note that $\lambda_i\approx 1$ and $\varepsilon\ll 1$ for reasonable survival rates.

In conclusion, even though there are extreme cases in which our methods might transform the noisy matrix into an invalid kernel matrix, we do not expect these problems to play any role because such extreme noise levels likely would render the \ac{QPU} output useless.

Note that the error bound in operator distance derived in Sec.~\ref{sec:sample_noise} is valid for the deviation of the statistical estimator from the noiseless kernel matrix $K$ or from the device-noisy kernel matrix $K^\mathrm{(dev)}$.
When applying post-processing methods however, this bound may not transfer to the output $K^\mathrm{(post)}$ in general.
Consequently, while being designed to counter device and finite sampling noise, the analytic error bound might become worse.

For $\regularization{THR}$ however, this bound is provably maintained \cite{guta2020fast}:
Splitting the indefinite matrix $\Bar{K}_M$ into the difference of two positive semi-definite matrices $K_+$ and $K_-$ with disjoint support, identifying $\regularization{THR}(\Bar{K}_M)=K_+$ and calculating the distance between the approximand\footnote{We here show the calculation when approximating $K$. It has to be replaced by $K^\mathrm{(dev)}$ accordingly when approximating the device-noisy matrix by sampling.} and $\Bar{K}_M$ yields
\begin{align}
    \Bar{K}_M \eqqcolon & K_+ - K_-\\ 
    \| K - \Bar{K}_M \|_\infty =& \| K - K_+ + K_- \|_\infty \\
    =&\max_{\|x\|_2=1} [x^T(K-K_+)^2 x \\
     & \quad+ \underset{\geq 0}{\underbrace{x^T (KK_- + K_-K+K_-^2)x}} ]\nonumber\\
    \geq& \max_{\|x\|_2=1} x^T(K-K_+)^2 x\\
    =& \| K-K_+ \|_\infty\\
    =& \| K-\regularization{THR}(\Bar{K}_M) \|_\infty
\end{align}
where we used the positive semi-definiteness of $K_-$ and that $K_\pm K_\mp = 0$ due to the disjoint support.

\subsection{Comparison of post-processing strategies}\label{sec:postprocessing_rating}

\begin{figure*}[ht]
    \centering
    \includegraphics[width=\textwidth]{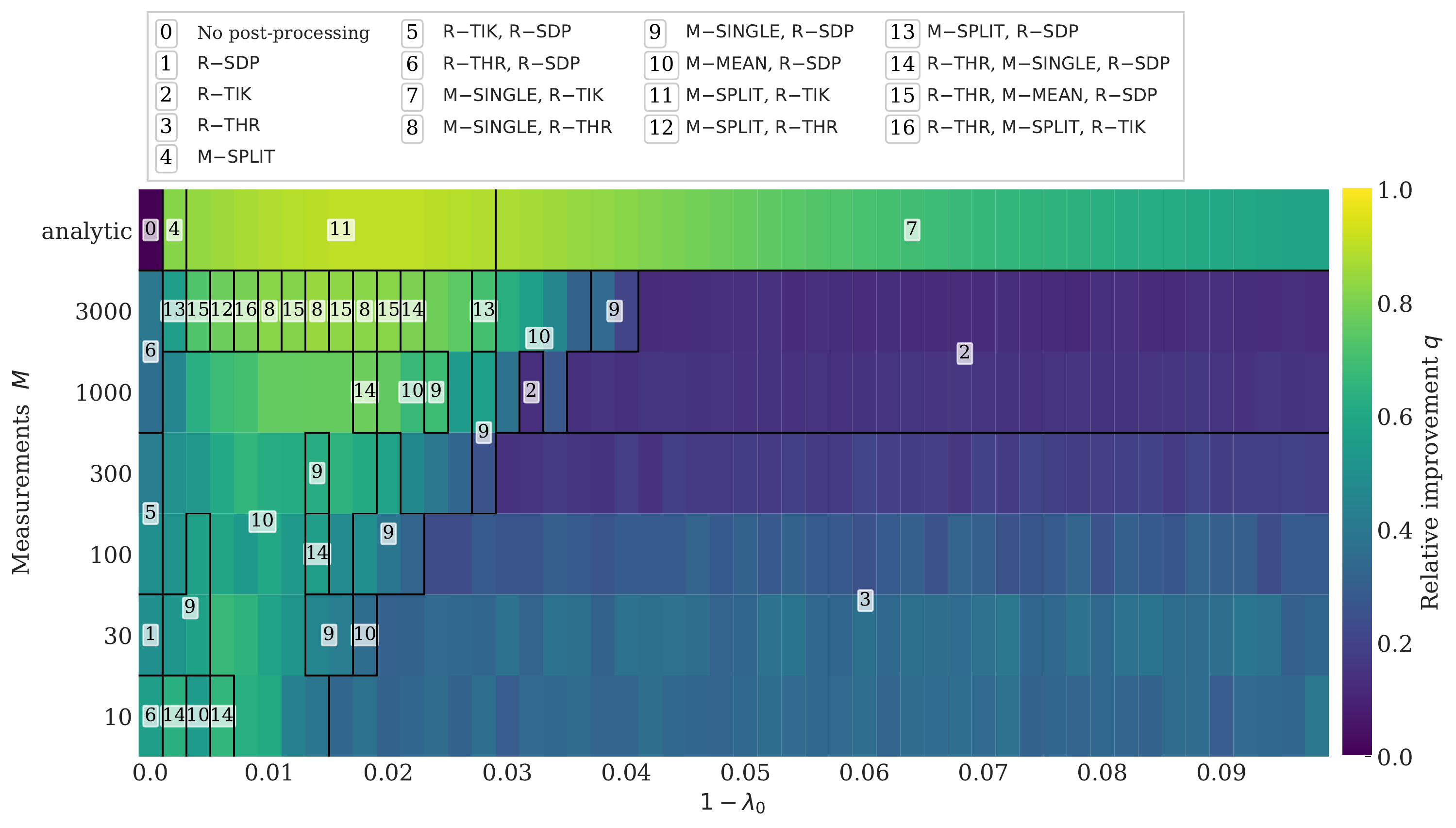}
    \caption{Relative improvement $q$ in alignment (see Eq.~\eqref{eq:def_quality_postproc}) between noisy and noiseless kernel matrices as in Fig.~\ref{fig:relative_improve_postprocessing} but for the best combination of post-processing steps \emph{per point}.
    We vary the base survival probability $\lambda_0$ of the depolarizing noise model (see Sec.~\ref{sec:noise_model}) and the number of circuit evaluations $M$ per kernel matrix entry, including the analytically exact limit $M\to\infty$ in the first row.
    For details on the noise mitigation and regularization methods see Sec.~\ref{sec:device_noise_mitigation} and~\ref{sec:regularization}, respectively.
    Black solid lines separate areas within which the same combination is best, the numbers label the combinations.
    }
    \label{fig:best_postprocessing}
\end{figure*}

There are many combinations of the post-processing techniques to choose from in order to counter both device noise and finite sampling noise.

First, we apply a regularization $\mathsf{R}_1$, including the option to not modify $K^\mathrm{(dev)}$ at all ($\opId$).
Second, we perform device noise mitigation $\mathsf{M}$ and third, we regularize again with $\mathsf{R}_2$.

For the two regularization steps $\mathsf{R}_{1,2}$, we may apply Tikhonov regularization ($\regularization{TIK}$), thresholding ($\regularization{THR}$) or the \ac{SDP} fixing the diagonal while thresholding ($\regularization{SDP}$), see Sec.~\ref{sec:regularization}.
For the mitigation step, we choose from single survival probability estimation based on a single ($\mitigation{SINGLE}$) or the mean ($\mitigation{MEAN}$) diagonal entry of $\Bar{K}_M$, and survival probability estimation per feature embedding ($\mitigation{SPLIT}$), see Sec.~\ref{sec:device_noise_mitigation}.

Naively, this yields $64$ combinations when including the trivial transformation $\opId$, out of which some are identical, e.g.\ $\opId, \opId, \mathsf{R}$ and $\mathsf{R}, \opId, \opId$.
In addition, there are special combinations in which methods effectively act like $\opId$:
First, combinations of the form $\regularization{SDP}, \mathsf{M}, \mathsf{R}_2$ for which $\mathsf{M}$ already receives a positive semi-definite input matrix with correct diagonal entries and thus will estimate the survival probability to be $1$.
Second, some combinations without mitigation (namely $\regularization{TIK}/\regularization{THR}, \opId, \regularization{TIK}/\regularization{THR}$) in which $\mathsf{R}_2$ would be redundant. Here we already excluded the combinations obeying the first pattern.
Excluding duplicates and these \enquote{reducible} combinations, we obtain $42$ reasonable, distinct strategies (including $\opId, \opId,\opId$) and for each of the outcomes $K^\mathrm{(post)}$ we compute the kernel alignment (see Eq.~\ref{eq:kernel_target_alignment_kernel_matrix}) with the noiseless matrix $K$.

In Fig.~\ref{fig:relative_improve_postprocessing} we only showed the best out of the resulting $42$ combinations, rated by the lowest alignment across all base survival probabilities and numbers of circuit evaluations.
This best combination is a single application of $\regularization{THR}$ (see Eq.~\eqref{eq:thresholding}), not making use of any mitigation or the second regularization step.
However, we note that the influence of sampling noise is rather large in the chosen domains of $M$ and $\lambda_0$, such that the rating by lowest achieved accuracy favors methods that are designed to counter sampling noise, such as $\regularization{THR}$.

Choosing the strategy in this way, our post-processing increases the alignment significantly (by up to $43.5\%$, in the regime of small $M$) and systematically (only negligible deterioration for $M\to\infty$), allowing for an improved estimate of the kernel matrix with fewer circuit evaluations.

In Fig.~\ref{fig:best_postprocessing} we evaluate the combinations of regularization and mitigation in more detail and it becomes apparent that the best choice depends on the noise regime.
We immediately see that in the domain of high noise (small numbers of circuit evaluations and lower survival probability, lower right) the result of our simple ranking in Sec.~\ref{ssec:numerics_mitigation} is confirmed and thresholding consistently is the best post-processing method (combination $3$).

When increasing $M$ while at the lower device noise level, Tikhonov regularization becomes more favorable than thresholding and still no device noise mitigation method is able to improve the matrix further (combination $2$).
This is remarkable because our combinations would allow for multiple processing steps to counter both, sampling and device noise.

For higher $\lambda_0$ we observe high variation in which combination is best, because many of them yield very similar alignments with the exact matrix so that statistical fluctuations become relevant.
However, for the majority of these datapoints with higher survival probabilities, the combination of single probability estimation based on a single or the mean of the matrix diagonal ($\mitigation{SINGLE}$/$\mitigation{MEAN}$) followed by \ac{SDP}-based regularization ($\regularization{SDP}$) is best (combinations $9$ and $10$).

Finally, for matrices without sampling noise our device noise mitigation techniques combined with Tikhonov regularization are the best choice, delivering far better results than simple regularization (combinations $5$ and $8$).
This indeed confirms that the high-level noise model of global depolarization, on which the mitigation techniques are based, is able to grasp some of the essential influence the device noise (of our more realistic depolarizing noise model) has.
Here the single probability estimation $\mitigation{SINGLE}$ seems to be better suited for lower survival probabilities and $\mitigation{SPLIT}$ for higher $\lambda_0$.

The complex appearance of Fig.~\ref{fig:best_postprocessing} underlines that the best choice of post-processing depends significantly on the noise level in the device and -- in relation to this level -- how many circuit evaluations are used for the kernel evaluation, so one should choose the method carefully.
The benefit of doing so compared to the first, simple rating in Sec.~\ref{ssec:numerics_mitigation} is an increase of the best achieved alignment improvement $q$ from $43.5\%$ to $84.6\%$ (or even to $90.2\%$ when considering the results without sampling noise in the first row of Fig.~\ref{fig:best_postprocessing}).

Finally, we note that the ranking of the post-processing combinations based on the alignment of $K^\mathrm{(post)}$ with the noiseless kernel matrix $K$ is not possible in applications that truely require the \ac{QPU}, as $K$ is not available in this case.
Instead one could evaluate the restored matrices based on their alignment with the ideal target matrix $K^\ast$.
We compared the ranking resulting from this idea with the one presented above and did not find any systematic relations between them, so that this application-oriented surrogate method to reconstruct $K$ may be discarded.
Whether it provides a better kernel matrix for classification is an open question, but if $K^\ast$ was optimal, there would not be any use in computing $K$ from the start.

We conclude that choosing the post-processing in practice remains challenging and requires a systematic analysis of both finite sampling and device noise effects on larger-scale kernel matrices. 

\section{Additional information on numerical experiments}
\label{app:numerical_experiments}

\subsection{Checkerboard}
    The checkerboard dataset is used for testing our error mitigation techniques for medium size kernels on a classical simulator, as well as for the noiseless simulation experiment. 
    The dataset consists of 30 training and 30 test datapoints, and was generated as follows.
    In the domain $[0,1]^2$ get a $4\times4$ grid with sites $i,j$ at coordinates $((2i+1)/8, (2j+1)/8)$ to prevent overlap between centroids and spilling out of the fixed domain.
    Next, we sampled points uniformly centered about each grid site.
    At the end, we assigned alternating classes to each of the sites, and finished by assigning all swarms of points the class corresponding to their centroid.
    
\subsection{Symmetric donuts}
    The symmetric donuts dataset ought to set the grounds for running a smaller kernel on actual quantum hardware.
    This artificial dataset has the same size as the previous one:  60 training and 60 test datapoints.
    The datapoints are generated by sampling points uniformly at random from a circle of radius $\sqrt{2}/2$ and then labeling them according to whether they fall within an inner circle of radius $1/2$ or without.
    We do this one time centering the circles on the $x$-axis, on the point $(1,0)$, giving the inner points label $1$ and the outer ones label $-1$.
    Next, we repeat the process for circles centered about the point $(-1,0)$ and this time exchange the labels: the inner point class is now $-1$ and the outer $+1$.
    This way we obtain a dataset contained in the domain~$[-(3+\sqrt{2})/2,(3+\sqrt{2})/2]\times[-\sqrt{2}/2,\sqrt{2}/2]$.

\subsection{Semi-artificial MNIST}
\begin{figure}
    \centering
    \includegraphics{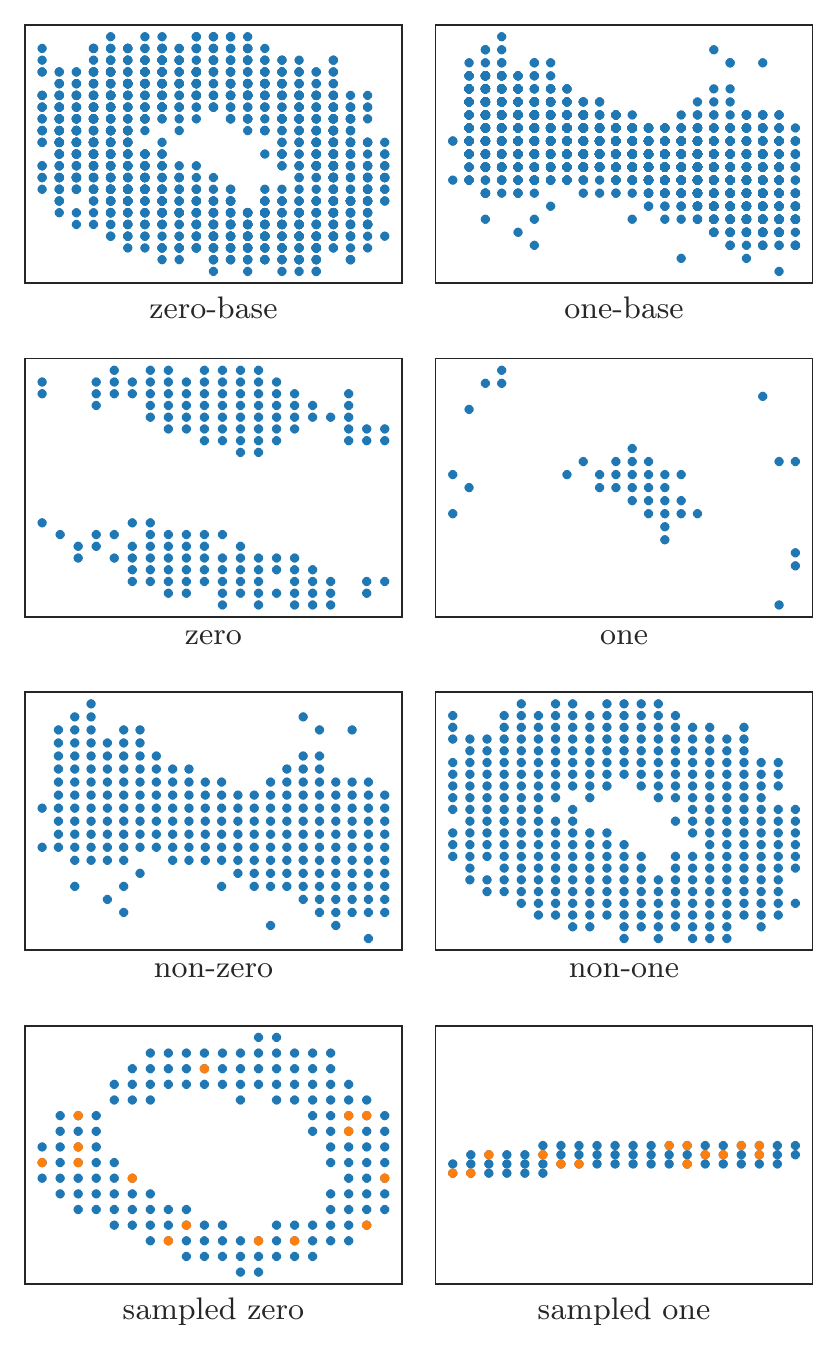}
    \caption{
    Various subsets of the MNIST dataset. All images appear rotated 90 degrees counter-clockwise.}
\label{fig:mnist_plots}
\end{figure}

    Moving our experiment closer to real-world data, we have also considered MNIST images. This dataset contains hand-written digits in images that have dimensions of 28x28 and a gray-scale from zero to 255, which we normalize to the range from zero to one. 
    To construct a semi-artificial dataset, we have randomly chosen $500$ images with labels of zero and one each, sampled three pixels with gray-scale values larger than 0.95 from each, and added all these coordinates to construct \enquote{zero-base} and \enquote{one-base} sets.
    We then constructed the dataset \enquote{zero} by selecting all coordinates in \enquote{zero-base} that are not contained in \enquote{one-base} and analogously for the dataset \enquote{one}. The dataset \enquote{non-zero} is a copy of dataset \enquote{one-base}, \enquote{non-one} is a copy of dataset \enquote{zero-base}. The datasets are shown in the top six plots of Fig.~\ref{fig:mnist_plots}.
\subsection{Ensemble MNIST}
    We previously used the semi-artificial MNIST dataset to train a classifier for classifying single pixel coordinates from a combined pool of $500$ images belonging to an image labeled zero (one) or not.
    We now use several of these classifiers in an ensemble, to classify \enquote{zero vs one} for individual images.
    This classification is done via the following steps: First, we sample $15$ pixels with a gray-scale value larger than $0.95$ from our image, and store them as coordinates. 
    This is shown in the bottom two plots of Fig.~\ref{fig:mnist_plots}.
    We then use the trained kernel classifiers to classify \enquote{zero} vs. \enquote{not zero} and \enquote{one} vs. \enquote{not one}. 
    Finally, we perform a majority vote for each coordinate.
    If the relative vote wins by less than two votes, we re-run our method with newly sampled points.

\subsection{Simulating device noise by depolarization}\label{sec:noise_model}
For the simulation of device noise in Sec.~\ref{ssec:numerics_mitigation} we use the following noise model:
After each unitary gate we apply single-qubit depolarizing noise channels $\calD_\lambda$ to each qubit the gate acted on (see Eq.~\eqref{eq:depolarizing_noise} with $N=1$).

Recalling the discussion in Sec.~\ref{sec:device_noise}, we remark that the qubit-wise depolarizing channel does commute with single-qubit but not with multi-qubit gates like the ring of controlled gates in our embedding circuit.
In this sense our model properly captures the case in which the device noise invalidates the adjoint approach, potentially destroying the positive semi-definiteness of the kernel matrix, and our post-processing strategies are challenged to correct this deviation.

The \emph{base survival probability} $\lambda_0$ quantifies the overall noise strength.
However, it is reasonable to expect that the noise strength for a specific gate on a \ac{QPU} depends on the duration of the pulses that implement the gate, leading to different effective noise levels for different embedded datapoints.
In order to capture this dependence, we rescale the base survival probability $\lambda_0$ for a rotation gate about the angle $\theta$ according to
\begin{align}
    \lambda=\left(1-\frac{\theta}{2\pi}\right)+\lambda_0\frac{\theta}{2\pi}
\end{align}
and fix the survival probability of the Hadamard and idling gate to $(1+\lambda_0)/2$ and $(1+3\lambda_0)/4$, respectively.

We do not simulate any device readout error explicitly but assume the presented implementation of depolarizing noise to represent the full device noise closely enough.
This assumption seems to be valid considering our results in Secs.~\ref{ssec:numerics_mitigation} and~\ref{ssec:numerics_hardware} and the accordance between them.

\section{CO\texorpdfstring{$_2$}{2} emission table}
\label{app:co2}
    \begin{center}
    \begin{tabular}[b]{l c}
    \hline
    \textbf{Numerical simulations} & \\
    \hline
    Total Kernel Hours [$\mathrm{h}$]& 7250\\
    Thermal Design Power Per Kernel [$\mathrm{W}$]& 4.6\\
    Total Energy Consumption Simulations [$\mathrm{kWh}$] & 32.8\\
    Average Emission Of CO$_2$ In Germany/USA [$\mathrm{kg/kWh}$]& 0.47\\
    Total CO$_2$-Emission For Numerical Simulations [$\mathrm{kg}$] & 15.5\\
    Estimated CO$_2$-Emission For \ac{QPU} usage [$\mathrm{kg}$] & 21.4 \\
    Were The Emissions Offset? & \textbf{Yes}\\
    \hline
    Total CO$_2$-Emission [$\mathrm{kg}$] & 36.9\\
    \hline
    \end{tabular}
    \end{center}

\end{document}

%% file: figures/QEK_block_tikz.tex
\begin{center}
\begin{tikzpicture}[scale=0.8, transform shape]
\begin{scope}
  \node[draw, fill=white] (gate0) at (0.5, 0) {$H$}; %Hadamard
  \node[draw, fill=white] (gate1) at (0.5, -1) {$H$}; %Hadamard
  \node[draw, fill=white] (gate2) at (0.5, -2) {$H$}; %Hadamard
  \node[draw, fill=white] (gate3) at (0.5, -3) {$H$}; %Hadamard
  \node[draw, fill=white] (gate4) at (0.5, -4) {$H$}; %Hadamard
  \node[draw, fill=white] (gate5) at (1.9, 0) {$RZ(x_1)$}; %RZ
  \node[draw, fill=white] (gate6) at (1.9, -1) {$RZ(x_2)$}; %RZ
  \node[draw, fill=white] (gate7) at (1.9, -2) {$RZ(x_1)$}; %RZ
  \node[draw, fill=white] (gate8) at (1.9, -3) {$RZ(x_2)$}; %RZ
  \node[draw, fill=white] (gate9) at (1.9, -4) {$RZ(x_1)$}; %RZ
  \node[draw, fill=white] (gate10) at (3.6999999999999997, 0) {$RY(\theta_{1})$}; %RY
  \node[draw, fill=white] (gate11) at (3.6999999999999997, -1) {$RY(\theta_{2})$}; %RY
  \node[draw, fill=white] (gate12) at (3.6999999999999997, -2) {$RY(\theta_{3})$}; %RY
  \node[draw, fill=white] (gate13) at (3.6999999999999997, -3) {$RY(\theta_{4})$}; %RY
  \node[draw, fill=white] (gate14) at (3.6999999999999997, -4) {$RY(\theta_{5})$}; %RY
  \draw[] (5.5, 0) -- +(0, -1) node[pos=0, circle, fill=black, inner sep=0pt,minimum size=3pt] {{}} node[pos=1, draw, fill=white] (gate15) {$RZ(\theta_{6})$}; %CRZ
  \draw[] (5.5, -2) -- +(0, -1) node[pos=0, circle, fill=black, inner sep=0pt,minimum size=3pt] {{}} node[pos=1, draw, fill=white] (gate16) {$RZ(\theta_{8})$}; %CRZ
  \draw[] (7.3, -1) -- +(0, -1) node[pos=0, circle, fill=black, inner sep=0pt,minimum size=3pt] {{}} node[pos=1, draw, fill=white] (gate17) {$RZ(\theta_{7})$}; %CRZ
  \draw[] (7.3, -3) -- +(0, -1) node[pos=0, circle, fill=black, inner sep=0pt,minimum size=3pt] {{}} node[pos=1, draw, fill=white] (gate18) {$RZ(\theta_{9})$}; %CRZ
  \draw[] (9.1, -4) -- +(0, 4) node[pos=0, circle, fill=black, inner sep=0pt,minimum size=3pt] {{}} node[pos=1, draw, fill=white] (gate19) {$RZ(\theta_{10})$}; %CRZ
\begin{pgfonlayer}{bg}
  \draw (0, 0) -- (10.0, 0);
  \draw (0, -1) -- (10.0, -1);
  \draw (0, -2) -- (10.0, -2);
  \draw (0, -3) -- (10.0, -3);
  \draw (0, -4) -- (10.0, -4);
\end{pgfonlayer}
\end{scope}
\end{tikzpicture}
\end{center}